\documentclass[12pt,a4paper]{article}
\usepackage{cite,mathtools,amsmath,amssymb}
\usepackage{graphicx}
\usepackage[usenames]{color}
\usepackage[bookmarksopen,colorlinks=true,linkcolor=dark_green,citecolor=dark_red,urlcolor=dark_red,linktocpage=false]{hyperref}

\definecolor{dark_blue}{rgb}{0,0,0.6}
\definecolor{dark_green}{rgb}{0,0.4,0}
\definecolor{dark_red}{rgb}{0.6,0,0}

\usepackage[height=22.5cm,width=16.5cm,centering]{geometry}

\def\thefootnote{\fnsymbol{footnote}}

\renewcommand{\thefootnote}{\fnsymbol{footnote}}
\begin{document}

\begin{titlepage}

\begin{center}

\hfill CTPU-17-28 \\
\hfill KEK-TH-1990 \\

\vskip .75in

{\fontsize{16pt}{0pt} \bf
Gravitational waves from first-order phase transitions:
\\ \vspace{5mm} 
}
{\fontsize{16pt}{0pt} \bf
Towards model separation by bubble nucleation rate 
}

\vskip .75in

{\large
Ryusuke Jinno$^{a}$, Sangjun Lee$^{b,a}$, 
\\ \vspace{3mm} 
Hyeonseok Seong$^{b}$ and Masahiro Takimoto$^{c,d}$
}

\vskip 0.25in

\begin{tabular}{ll}
$^{a}$ &\!\! {\em Center for Theoretical Physics of the Universe, Institute for Basic Science (IBS),}\\
&{\em Daejeon 34051, Korea}\\[.3em]
$^{b}$ &\!\! {\em Department of Physics, KAIST, Daejeon 34141, Korea,}\\[.3em]
$^{c}$ &\!\! {\em Department of Particle Physics and Astrophysics, Weizmann Institute of Science,}\\
&{\em Rehovot 7610001, Israel}\\[.3em]
$^{d}$ &\!\! {\em Theory Center, High Energy Accelerator Research Organization (KEK),}\\
&{\em Oho, Tsukuba, Ibaraki 305-0801, Japan}\\[.3em]
\end{tabular}

\end{center}
\vskip .5in

\begin{abstract}
We study gravitational-wave production from bubble collisions in a cosmic first-order phase transition,
focusing on the possibility of model separation by the bubble nucleation rate dependence of the resulting 
gravitational-wave spectrum.
By using the method of relating the spectrum with the two-point correlator 
of the energy-momentum tensor $\left< T(x)T(y) \right>$,
we first write down analytic expressions for the spectrum 
with a Gaussian correction to the commonly used nucleation rate, 
$\Gamma \propto e^{\beta t}\rightarrow e^{\beta t-\gamma^2t^2}$,
under the thin-wall and envelope approximations.
Then we quantitatively investigate how the spectrum changes with the size of the Gaussian correction.
It is found that the spectral shape shows ${\mathcal O}(10)\%$ deviation 
from $\Gamma \propto e^{\beta t}$ case for some physically motivated scenarios.
We also briefly discuss detector sensitivities required to distinguish different spectral shapes.
\end{abstract}

\end{titlepage}

\tableofcontents
\thispagestyle{empty}

\renewcommand{\thepage}{\arabic{page}}
\setcounter{page}{1}
\renewcommand{\thefootnote}{$\diamondsuit$\arabic{footnote}}
\setcounter{footnote}{0}

\newpage
\setcounter{page}{1}

\section{Introduction}
\label{sec:Intro}
\setcounter{equation}{0}

Gravitational waves (GWs) offer the exciting possibility of 
probing the early Universe well before the Big Bang Nucleosynthesis
and of revealing unknown high-energy particle physics.
Possible cosmological sources of GWs include 
inflationary quantum fluctuations~\cite{Starobinsky:1979ty}, 
preheating~\cite{Khlebnikov:1997di},
topological defects~\cite{Vilenkin:2000jqa}
and first-order phase transitions~\cite{Witten:1984rs,Hogan:1986qda},
and GWs from these sources are expected to be detected in the near future.
In fact, ground-based GW detectors such as advanced LIGO~\cite{TheLIGOScientific:2014jea}, 
KAGRA~\cite{Somiya:2011np} and VIRGO~\cite{TheVirgo:2014hva} are now in operation,
and the first detections of GWs from black hole binaries 
by advanced LIGO collaboration~\cite{Abbott:2016blz,Abbott:2016nmj,Abbott:2017vtc}
have opened up a new era of GW astronomy.
In the future, space-borne detectors such as LISA~\cite{Seoane:2013qna}, 
DECIGO~\cite{Seto:2001qf} and BBO~\cite{Harry:2006fi} are expected to start GW cosmology.

Among various sources of GWs in the early Universe, 
we focus on first-order phase transitions~\cite{Witten:1984rs,Hogan:1986qda} in this paper.
Though first-order phase transitions do not occur 
in the standard model~\cite{Kajantie:1996mn,Gurtler:1997hr,Csikor:1998eu},
there are various models which predict first-order phase transitions 
(see {\it e.g.} Refs.~\cite{Models,Jinno:2016knw,Kobakhidze:2017mru,Iso:2017uuu,Cai:2017tmh},
and also Refs.~\cite{Caprini:2015zlo,Cai:2017cbj} and references therein for reviews).
Furthermore, planned detectors are sensitive to the transition dynamics around TeV-PeV scales,
and such GWs provide an opportunity of probing new physics 
beyond the standard model.

In thermal first-order phase transitions,
bubbles of the true vacuum nucleate at some stage in the history of the Universe.
They then expand because of the pressure difference between the true and false vacua,
and the transition completes after they eventually collide with each other.\footnote{
Though the Universe is covered with the true vacuum region when the bubbles collide with each other,
it does not mean that the GW production ceases by this time. See below (``sound waves").
}
Gravitational waves are produced during this process 
through their coupling to the energy-momentum tensor of the system.
In other words, various properties of the energy-momentum tensor 
around the time of the high-energy transition 
are imprinted on the resulting GW spectrum.
From the viewpoint of both 
knowing the dynamics in the early Universe and identifying the underlying particle physics,
it is of great importance to study what kind of properties one may extract from the GW spectrum.

In first-order phase transitions, 
the main ingredients which determine the behavior of the energy-momentum tensor are classified as
\begin{itemize}
\item[(1)]
Spacetime distribution of bubbles 
({\it i.e.}, nucleation rate of bubbles),
\item[(2)]
Energy-momentum tensor profile around a bubble wall,
\item[(3)]
Dynamics after bubble collisions.
\end{itemize}
Much effort has been made to reveal the effects of (2) and (3) on the GW spectrum.
For example, 
in the first numerical simulation of GW production in a vacuum transition
({\it i.e.}, a system with only the scalar field which drives the transition)~\cite{Kosowsky:1991ua},
it was found that the sourcing process is almost free from the detailed structure of the bubble walls.
It was also found that the main GW production comes from the uncollided bubble walls.
These findings led to ``thin-wall" and ``envelope" approximations,
which were frequently used in subsequent works~\cite{Kosowsky:1992rz,Kosowsky:1992vn,Kamionkowski:1993fg}:
the former corresponds to assuming 
infinitely thin concentration of energy and momentum for (2), 
while the latter corresponds to assuming
an instant damping of the wall energy and momentum for (3).
Also, the authors of Refs.~\cite{Hindmarsh:2013xza,Hindmarsh:2015qta,Hindmarsh:2017gnf}
have recently pointed out in a series of numerical simulations
that in scalar-fluid systems the bulk motion of the fluid works as a long-lasting source for GWs (``sound waves")
even after bubbles collide with each other.\footnote{
Turbulence is another important source for GWs 
in first-order phase transitions~\cite{Kamionkowski:1993fg,Kosowsky:2001xp,Nicolis:2003tg,
Caprini:2006jb,Caprini:2009yp,Kahniashvili:2009mf}.
}
This discovery has significantly changed our understanding on (3).

On the other hand, there are much less studies on the possibility of extracting information on (1) 
from the GW spectrum.
However, in this paper we stress the importance of such studies because 
the spacetime distribution of bubbles is determined by the underlying 
particle physics through the time evolution of the bubble nucleation rate.
In the literature
the bubble nucleation rate in thermal first-order phase transitions has often 
been approximated simply by an exponential form $\Gamma \propto e^{\beta t}$.
From the viewpoint of extracting 
as much information on the underlying particle physics as possible, 
we investigate the effect of the nucleation rate on the GW spectrum
by going beyond the linear approximation $\Gamma \propto e^{\beta t}$.\footnote{
It has been pointed out that deviations from the conventional nucleation rate
$\Gamma \propto e^{\beta t}$ are realized 
in some models~\cite{Espinosa:2008kw,Huber:2015znp,Leitao:2015fmj,Megevand:2016lpr,Kobakhidze:2017mru,Cai:2017tmh}
(see also Appendix~\ref{app:Model}).
}\footnote{
Though in Ref.~\cite{Weir:2016tov} it has been reported that simultaneous nucleation of bubbles
changes the spectral peak frequency and amplitude from 
those with the exponential nucleation rate by some factor,
such information on the peak would be mixed up with other parameter dependences.
In contrast we investigate how much information the {\it spectral shape itself} contains.
}

For this purpose we adopt the method of relating the GW spectrum 
with the two-point ensemble average of the energy-momentum tensor $\left< T(x)T(y) \right>$.
This method, which utilizes the stochastic nature of produced GWs, 
was first used in Ref.~\cite{Caprini:2007xq} in the context of bubble dynamics in first-order phase transitions.
In Ref.~\cite{Jinno:2016vai} it has been pointed out that 
under the thin-wall approximation various contributions to this two-point correlator reduce to only two classes
and the resulting spectrum becomes analytically calculable.
As a result, the GW spectrum by the numerical simulation with the same setup in Ref.~\cite{Huber:2008hg},
{\it i.e.}, the one with the thin-wall and envelope approximations, 
was derived analytically in Ref.~\cite{Jinno:2016vai}.
This direction of study has recently been extended in Ref.~\cite{Jinno:2017fby},
and a general form of the GW spectrum without the envelope approximation has been given and analyzed.
In this paper, we focus on the effect of (1) on the GW spectrum.
Therefore we adopt the simplest setup for (2) and (3): the thin-wall and envelope approximations.
Though these approximations may not give a satisfactory description of the system,\footnote{
In fact, it has been pointed out in Ref.~\cite{Weir:2016tov} that 
the modeling of the system with these approximations does not hold good
when the bubble walls reach a low terminal velocity 
(in the sense that the gamma factor $\gamma_w \equiv 1/\sqrt{1 - v_w^2}$ 
of the wall velocity $v_w$ satisfies $\gamma_w \lesssim {\mathcal O}(1)$)
because of the long-lasting nature of the bulk motion of the fluid as GW sources.
}
our study will give to some extent a quantitative measure
of the dependence of the spectral shape on the nucleation rate.\footnote{
It would be possible to extend the present setup 
and remove the envelope approximation by using the results of Ref.~\cite{Jinno:2017fby}.
We leave such a study to future work.
}

The organization of the paper is as follows.
In Sec.~\ref{sec:Formalism} we first make clear our assumptions and approximations
on (1)--(3) above, and then present the formalism to calculate the GW spectrum.
In Sec.~\ref{sec:Analytic} we give analytic expressions for the spectrum.
In Sec.~\ref{sec:Numerical} we evaluate the expressions with numerical methods.
Sec.~\ref{sec:Conclusion} is devoted to conclusions.
We also discuss typical models which realize deviations in the nucleation rate 
from the exponential form in Appendix~\ref{app:Model},
and present a detailed derivation of the analytic expression of the spectrum 
in Appendix~\ref{app:Derivation}. 
We briefly examine the asymptotic behavior of the spectrum in the limit of small Gaussian correction 
in Appendix~\ref{app:Asymptotic}.

\section{Formalism}
\label{sec:Formalism}
\setcounter{equation}{0}

In this section we first summarize the assumptions and approximations adopted in this paper,
and then explain the method of relating the GW spectrum to the two-point correlator of 
the energy-momentum tensor $\left< T(x)T(y) \right>$.

\subsection{Assumptions and approximations}
\label{subsec:AA}

\subsubsection*{Thin-wall and envelope approximations}

First, we introduce two important approximations which determine (2) and (3):
the thin-wall and envelope approximations.
The former assumes that the energy released from the transition is 
localized around the thin surfaces of bubbles.
We parameterize the energy-momentum tensor of an uncollided bubble as
\begin{align}
T_{Bij}(x)
&= \rho_B(x)\widehat{(x - x_n)}_i\widehat{(x - x_n)}_j,
\label{eq:TB}
\end{align}
where $x = (t_x, \vec{x})$ denotes a spacetime point 
and $x_n = (t_{xn}, \vec{x}_n)$ is the nucleation point of the bubble.
Also, $\hat{\bullet}$ denotes the unit vector in $\mathbf{\bullet}$ direction.
In addition, we take $\rho_B$ to be
\begin{align}
\rho_B(x)
&=
\left\{
\begin{array}{cc}
\displaystyle 
\frac{4\pi}{3} r_B(t_x,t_{xn})^3 \kappa\rho_0
\Big/ 4\pi r_B(t_x,t_{xn})^2 l_B
&
r_B(t_x,t_{xn}) < |\vec{x} - \vec{x}_n| < r'_B(t_x,t_{xn}) \\
0
& 
{\rm otherwise}
\end{array}
\right. ,
\label{eq:rhoB}
\end{align}
where $r_B, r'_B$ are the inner and outer radii of the bubble
\begin{align}
r_B(t_x,t_{xn})
&= 
v(t_x - t_{xn}), 
\;\;\;
r'_B(t_x,t_{xn})
= r_B(t_x,t_{xn}) + l_B.
\end{align}
Here $l_B$ denotes the thickness of the wall, and we take $l_B \to 0$ in the final step. 
Also, $\rho_0$ is the released energy density, 
and the efficiency factor $\kappa$ determines the fraction of $\rho_0$ 
transformed into the macroscopic energy around the wall.\footnote{
The efficiency factor $\kappa$ can be calculated by following Ref.~\cite{Espinosa:2010hh}.
}
In addition, $v$ is the wall velocity, which we assume to be constant throughout the paper.
Note that the modeling (\ref{eq:rhoB}) takes into account 
the proportionality of the released energy to the volume of the bubble
and its localization within the bubble wall with width $l_B$.
In the following calculations, we neglect the effect of cosmic expansion.

\subsubsection*{Nucleation rate}
\label{sec:nucr}

In this paper we parameterize the bubble nucleation as
\begin{align}
\Gamma(t)
&= \Gamma_*e^{\beta t - \gamma^2 t^2},
\label{eq:Gamma}
\end{align}
where $\beta$ and $\gamma$ are some constants 
and $\Gamma_*$ is the nucleation rate at some reference time, which we take to be $t = 0$. 
Though $\gamma$ is often set to be zero in the literature,
we aim to investigate the effect of this Gaussian correction on the spectral shape in this paper.

\begin{figure}[t!]
\begin{center}
\includegraphics[width=0.55\columnwidth]{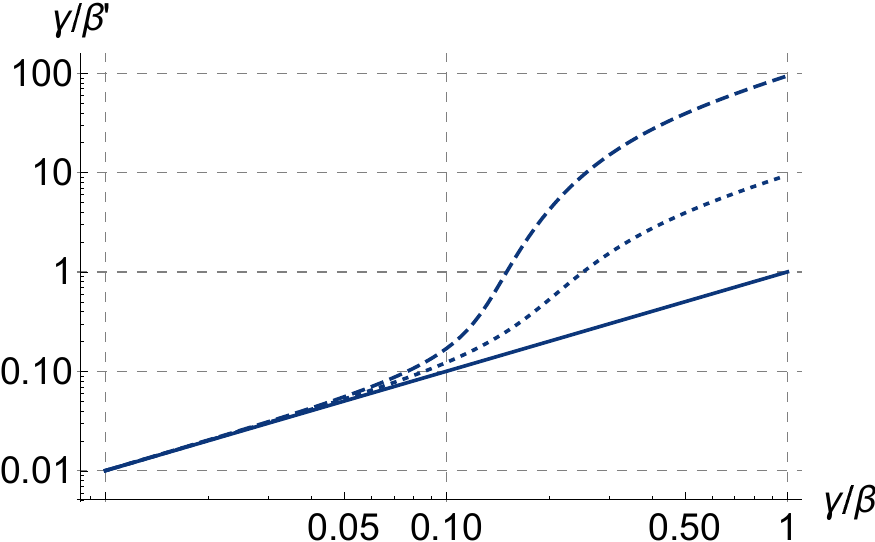} 
\caption{\small
Relation (\ref{eq:rel}) between dimensionless combinations 
$\gamma/\beta$ and $\gamma/\beta'$.
In this plot $\beta/H_*$ is taken to be $1, 10, 100$ from bottom to top.
}
\label{fig:GammaGammapr}
\end{center}
\end{figure}

In the literature, the origin of time $t = 0$ is often defined by the condition
\begin{align}
\Gamma_* 
&= H_*^4,
\end{align}
where $H_*$ is the Hubble parameter at the time of transition.
Note that in this paper we neglect the effect of cosmic expansion and thus $H_*$ is an input parameter.
Though this definition completely specifies the form of the nucleation rate (\ref{eq:Gamma}),
it leads to redundancy among the parameters $(H_*, \beta, \gamma)$ 
in presenting the GW spectrum.
This is because of a time-shift invariance:
$\Gamma(t) = \Gamma_*e^{\beta t - \gamma^2 t^2}$ and 
$\Gamma(t) = \Gamma_*e^{\beta (t + \Delta t) - \gamma^2 (t + \Delta t)^2}$
give the same GW spectrum for an arbitrary $\Delta t$
because the GW spectrum is obtained after integrating over the whole period of time.
Therefore, in presenting the final results in Sec.~\ref{sec:Numerical},
we eliminate this redundancy by choosing $\Delta t$ so that the nucleation rate is 
parameterized as
\begin{align}
\Gamma(t)
&= \Gamma_*' e^{\beta' t' - \gamma^2 t'^2},
\;\;\;\;
\Gamma_*' 
= \beta'^4,
\;\;\;\;
t
= t' + \Delta t.
\label{eq:Gammapr}
\end{align}
These new parameters satisfy\footnote{
Substituting $t = t' + \Delta t$, we have
\begin{align}
\Gamma_* e^{\beta (t' + \Delta t) - \gamma^2 (t' + \Delta t)^2}
&= 
\Gamma_*' e^{\beta' t' - \gamma^2 t'^2},
\end{align}
and we can read off the relations 
\begin{align}
\Gamma_*'
&= 
\Gamma_* e^{\beta \Delta t - \gamma^2 \Delta t^2},
\;\;\;\;
\beta' 
= \beta - 2\gamma^2 \Delta t.
\end{align}
The time shift $\Delta t$ is determined by the requirement $\Gamma_*' = \beta'^4$.
Note that it is always possible to find such $\Delta t$.
Also, eliminating $\Delta t$ from this relation, we obtain Eq.~(\ref{eq:rel}).
}
\begin{align}
\frac{H_*^4}{\gamma^4} e^{\frac{\beta^2}{4\gamma^2}}
&= 
\frac{\beta'^4}{\gamma^4} e^{\frac{\beta'^2}{4\gamma^2}},
\label{eq:rel}
\end{align}
which should be regarded as an equation to determine $\gamma/\beta'$ from the old parameters
$\beta/H_*$ and $\gamma/\beta$.
This new parameterization has a physical interpretation that 
$t' = 0$ corresponds to a typical transition time
for $\beta'/H_* \gg 1$ and $\gamma/\beta' \ll 1$.\footnote{
We assume that higher order corrections such as $t'^n$ ($n \geq 3$) 
in the exponent can be neglected:
this is a natural assumption as long as $\gamma/\beta' \ll 1$ holds.
}
Also, noting that the spectral shape is determined only by dimensionless quantities,
we see that only $\gamma/\beta'$ determines the spectral shape.
In Fig.~\ref{fig:GammaGammapr} we show the relation between $\gamma/\beta'$ and $\gamma/\beta$ 
for several $\beta / H_*$.

In Table~\ref{tbl:Param} we summarize several parameterizations of the nucleation rate obtained 
by the time shift.
Parameterization 1 corresponds to the original one (\ref{eq:Gamma}),
while Parameterization 2 corresponds to the one introduced in Eq.~(\ref{eq:Gammapr}).
Also, we use Parameterization 3 when we introduce analytic expressions for the spectrum
in Sec.~\ref{sec:Analytic} because it makes the expressions simplest.
In presenting the final results in Sec.~\ref{sec:Numerical}, we use Parameterization 2.

Finally, we mention typical values of $\beta'/H_*$ and $\gamma/\beta'$.
The parameter $\beta'/H_*$ largely depends on the particle physics setup
and typically varies within $\sim \mathcal{O}{(10^{1-5})}$.
On the other hand, the dependence of $\gamma/\beta'$ on setups is relatively mild
and it takes $\sim \mathcal{O}(0.1)$,
though larger values are possible in very strong phase transitions~\cite{Espinosa:2008kw,Huber:2015znp,Leitao:2015fmj,Megevand:2016lpr,Kobakhidze:2017mru,Cai:2017tmh}.
In Appendix~\ref{app:Model}, 
we show that the typical value of $\gamma/\beta'$ is $\mathcal{O}(0.1)$,
and calculate it for some motivated models.
Therefore, note that if we have $\mathcal{O}(0.1)$ sensitivity on $\gamma/\beta'$ in future observations
we have the possibility of distinguishing models by studying the spectral shape of GWs.

\begin{table}
\begin{center}
\caption{
Equivalent parameterizations of the nucleation rate used in the paper.
}
\vskip 0.1in
\renewcommand{\arraystretch}{1.2}
\begin{tabular}{|c||c|c|c|} \hline 
Parameterization & 1 & 2 & 3 \\ \hline\hline
Variables 
& $(\Gamma_*, \beta, \gamma)$ 
& $(\beta', \gamma)$ 
& $(\Gamma_*'', \gamma)$ \\ \hline
Nucleation rate $\Gamma$ 
& $\Gamma_* e^{\beta t - \gamma^2 t^2}$ 
& $\Gamma_*' e^{\beta' t' - \gamma^2 t'^2}$, $\Gamma_*' = \beta'^4$
& $\Gamma_*'' e^{- \gamma^2 t''^2}$ \\ \hline
Section 
& Sec.~\ref{sec:Formalism} 
& Sec.~\ref{sec:Formalism}, \ref{sec:Analytic}  
& Sec.~\ref{sec:Analytic}, Appendix~\ref{app:Derivation} \\ \hline
\end{tabular}
\renewcommand{\arraystretch}{1} 
\label{tbl:Param}
\end{center}
\end{table}

\subsection{GW spectrum as energy-momentum tensor correlation}
\label{sec:GWcorr}

\subsubsection*{GW spectrum around the time of transition}

As mentioned above, we neglect the effect of cosmic expansion during the transition.
Under this approximation the metric is well described by 
\begin{align}
ds^2
&= - dt^2 + (\delta_{ij}+2h_{ij})dx^idx^j.
\end{align}
The evolution equation for the tensor perturbations $h_{ij}$ 
(satisfying the transverse and traceless conditions $h_{ii} = \partial_i h_{ij} = 0$) is given by
\begin{align}
\ddot{h}_{ij}(t,\vec{k})+k^2h_{ij} (t,\vec{k}) 
&= {8\pi G } \Pi_{ij}(t,\vec{k}),
\label{eq:hEOM}
\end{align}
where the dot denotes the time derivative
and we have moved to the Fourier space labeled by the three momentum $\vec{k}$.
The source term $\Pi_{ij}$  denotes the projected energy-momentum tensor:
\begin{align}
\Pi_{ij}(t,\vec{k})
&= K_{ijkl}(\hat{k}) T_{kl}(t,\vec{k}),
\label{eq:PiKT}
\\
K_{ijkl}(\hat{k})
&\equiv P_{ik}(\hat{k})P_{jl}(\hat{k}) - \frac{1}{2}P_{ij}(\hat{k})P_{kl}(\hat{k}), 
\;\;\;\;
P_{ij}(\hat{k})
\equiv \delta_{ij}-\hat{k}_i\hat{k}_j.
\label{eq:K}
\end{align}
We assume that the source is switched on from $t_{\rm start}$ to $t_{\rm end}$,
which we take $t_{\rm start/end} \rightarrow \mp \infty$ in the following calculation.

The energy density of GWs is given by
\begin{align}
\rho_{\rm GW}(t)
&= \frac{\langle \dot{h}_{ij}(t,\vec{x})\dot{h}_{ij}(t,\vec{x}) \rangle}{8\pi G},
\label{eq:rhoGW}
\end{align}
where the angular bracket denotes 
taking both an oscillation average for several oscillation periods and an ensemble average.
In terms of the Fourier mode, the energy density per each logarithmic wavenumber 
is expressed as
\begin{align}
\Omega_{\rm GW}(t,k) 
&\equiv \frac{1}{\rho_{\rm tot}(t)} \frac{d\rho_{\rm GW}}{d\ln k}(t,k)
=
\frac{k^3}{16\pi^3 G}P_{\dot{h}}(t,k).
\label{eq:OmegaGW}
\end{align}
Here we have normalized the energy density by the total energy density $\rho_{\rm tot}$ at time $t$ 
to define $\Omega_{\rm GW}$.
Also, we have defined the power spectrum $P_{\dot{h}}$ as
\begin{align}
\langle \dot{h}_{ij}(t,\vec{k}) \dot{h}_{ij}^*(t,\vec{q}) \rangle
&= (2\pi)^3 \delta^{(3)}(\vec{k} - \vec{q}) P_{\dot{h}}(t,k).
\label{eq:dothdoth}
\end{align}
Since $h_{ij}$ is related to the source term $\Pi$ through Eq.~(\ref{eq:hEOM}),
we can express $\Omega_{\rm GW}$ in terms of $\Pi$.
For this purpose let us define the unequal-time correlator of the source term as
\begin{align}
\langle \Pi_{ij}(t_x,\vec{k})\Pi^*_{ij}(t_y,\vec{q}) \rangle
&= (2\pi)^3 \delta^{(3)}(\vec{k} - \vec{q})\Pi(t_x,t_y,k).
\label{eq:PiPi}
\end{align}
This correlator is related to the original energy-momentum tensor through
\begin{align}
\Pi(t_x,t_y,k) 
&= K_{ijkl}(\hat{k})K_{ijmn}(\hat{k})
\int d^3r \; e^{i \vec{k} \cdot \vec{r}} \langle T_{kl} T_{mn} \rangle (t_x,t_y,\vec{r}).
\label{eq:Pi}
\end{align}
Here the quantity in the integrand is defined as
\begin{align}
\langle T_{ij} T_{kl} \rangle (t_x,t_y,\vec{r})
&\equiv \langle T_{ij}(t_x,\vec{x}) T_{kl}(t_y,\vec{y}) \rangle,
\end{align}
with $\vec{r} \equiv \vec{x} - \vec{y}$.
Note that the spacial homogeneity of the system makes 
the correlator depend on $\vec{x}$ and $\vec{y}$ only through the combination $\vec{r}$.
Now, using the Green function method, we obtain from Eq.~(\ref{eq:OmegaGW})
(see Refs.~\cite{Caprini:2007xq,Jinno:2016vai,Jinno:2017fby})
\begin{align}
\Omega_{\rm GW} (t,k)
&= \frac{2Gk^3}{\pi \rho_{\rm tot}(t)}
\int_{t_{\rm start}}^{t_{\rm end}} dt_x
\int_{t_{\rm start}}^{t_{\rm end}} dt_y \;
\cos(k(t_x - t_y))\Pi (t_x,t_y,k),
\;\;\;\;
t > t_{\rm end},
\label{eq:OmegaPi}
\end{align}
where we have assumed that the source term exists only from $t_{\rm start}$ to $t_{\rm end}$.
This equation allows us to obtain the GW spectrum straightforwardly once we find expressions for $\Pi(t_x,t_y,k)$,
or equivalently the two-point correlator $\langle T(x)T(y) \rangle$.
Note that, though we have put the argument $t$,
the L.H.S. essentially does not depend on it as long as $t > t_{\rm end}$ 
because there is no production or dilution of GWs once the source term is switched off.

We may further factor out some parameter dependences from Eq.~(\ref{eq:OmegaPi}).
We first define the fraction of the released energy density 
to the background radiation energy density $\rho_{\rm rad}$ just before the transition as
\begin{align}
\alpha
&\equiv \frac{\rho_0}{\rho_{\rm rad}},
\;\;\;\;\;\;
\rho_{\rm tot}
= \rho_0 + \rho_{\rm rad}.
\end{align}
Then, defining the dimensionless power spectrum $\Delta$ as
\begin{align}
\Omega_{\rm GW}(t,k)
&\equiv
\kappa^2 \left(\frac{H_*}{\beta}\right)^2\left(\frac{\alpha}{1+\alpha}\right)^2
\Delta(k/\beta),
\label{eq:OmegaDelta}
\end{align}
we can factor out $G$, $\kappa$, $\rho_0$ and $\rho_{\rm tot}$ dependences.\footnote{
This is understood as follows.
Note that the factor in front of $\Delta$ in Eq.~(\ref{eq:OmegaDelta})
is proportional to $(G \kappa \rho_0)^2 / G / \rho_{\rm tot}$.
The combination $G \kappa \rho_0$ comes from 
the dependence of the source term in Eq.~(\ref{eq:hEOM}) on this quantity.
Then the GW energy density (\ref{eq:rhoGW}) gives the dependence $(G \kappa \rho_0)^2 / G$.
Finally normalization by $\rho_{\rm tot}$ in the definition of $\Omega_{\rm GW}$ in Eq.~(\ref{eq:OmegaGW}) 
gives the above dependence.
}
The new power spectrum $\Delta$ is expressed in terms of $\Pi$ as
\begin{align}
\Delta(k/\beta)
&= 
\frac{3}{4\pi^2}\frac{\beta^2k^3}{\kappa^2\rho_0^2}
\int_{t_{\rm start}}^{t_{\rm end}} dt_x
\int_{t_{\rm start}}^{t_{\rm end}} dt_y \;
\cos(k(t_x - t_y))\Pi (t_x,t_y,k),
\label{eq:DeltaPi}
\end{align}
where we have used the Friedmann equation $H_*^2 = (8\pi G/3)\rho_{\rm tot}$.
The spectrum $\Delta$ depends on dimensionless combinations 
such as $k/\beta$, $\gamma/\beta$ and $v$.
Here we have kept only $k/\beta$ in the argument in order to make clear that $\Delta$ is a spectrum.

\subsubsection*{GW spectrum at present}

Gravitational waves are redshifted after production until the present time.
The present frequency $f_0$ and amplitude $\Omega_{{\rm GW},0}$ are obtained by taking into account that
GWs behave as non-interacting radiation well inside the horizon
(see Refs.~\cite{Jinno:2016vai,Jinno:2017fby}):
\begin{align}
f_0
&= 
1.65 \times 10^{-5} {\rm Hz}
\left( \frac{f_*}{\beta} \right) \left( \frac{\beta}{H_*} \right)
\left( \frac{T_*}{10^2{\rm GeV}} \right)
\left( \frac{g_*}{100} \right)^{\frac{1}{6}},
\label{eq:fPresent}
\\
\Omega_{{\rm GW},0} (f_0) h^2 
&=
1.67\times 10^{-5}\kappa^2 \Delta (k/\beta)
\left( \frac{H_*}{\beta} \right)^2
\left( \frac{\alpha}{1 + \alpha} \right)^2
\left( \frac{g_*}{100} \right)^{-\frac{1}{3}},
\label{eq:OmegaPresent}
\end{align}
where $g_*$ is the total number of relativistic degrees of freedom in the thermal bath at temperature $T_*$, 
and $f_*$ is the frequency at the time of transition.\footnote{
In this paper we use $k$ to denote the physical wavenumber
at the transition time.
}
In Eq.~(\ref{eq:OmegaPresent}),
the present frequency $f_0$ is related to 
the wavenumber at the transition time $k$ through the relation (\ref{eq:fPresent})
with $f_* = k/2\pi$.
Note that the spectral shape is encoded in $\Delta$.

\section{Analytic expressions}
\label{sec:Analytic}
\setcounter{equation}{0}

\subsection{Basic strategy}

We briefly summarize the basic strategy to calculate the GW spectrum.
This subsection is essentially the same as Ref.~\cite{Jinno:2017fby}.
From Eq.~(\ref{eq:DeltaPi}), 
we see that the spectrum is calculated from the unequal-time correlator $\Pi(t_x,t_y,k)$,
which is the Fourier transform of $\Pi(t_x,t_y,\vec{r})$.
Since the correlator $\Pi(t_x,t_y,\vec{r})$ is equivalent to $\left< T(t_x,\vec{x}) T(t_y,\vec{y}) \right>$
with $T$ symbolically denoting the energy-momentum tensor,
the procedure we have to follow to obtain the spectrum is
\begin{itemize}
\item
Fix the spacetime points $x = (t_x, \vec{x})$ and $y = (t_y, \vec{y})$.
\item
Find bubble configurations giving nonzero $T(x)T(y)$, 
calculate the probability for each configuration to occur,
and estimate the value of $T(x)T(y)$ in each case.
\item
Sum over all possible configurations.
\end{itemize}
As in Refs.~\cite{Jinno:2016vai,Jinno:2017fby},
one may classify the bubble configurations 
depending on whether the contributions to $T(t_x,\vec{x})$ and $T(t_y,\vec{y})$ 
come from the same bubble or different bubbles.
This consideration leads to the following classification:
\begin{itemize}
\item
Single-bubble spectrum $\Delta^{(s)}$,
\item
Double-bubble spectrum $\Delta^{(d)}$.
\end{itemize}
The final spectrum $\Delta$ becomes the sum of the two:
$\Delta = \Delta^{(s)} + \Delta^{(d)}$.
Note that the single-bubble spectrum does not mean contributions from an isolated bubble (which would vanish)
but it means that one sums over the configurations in which the wall fragments affecting $x$ and $y$
come from the same nucleation point (see Appendix~H of Ref.~\cite{Jinno:2017fby} for more explanation).

\subsection{Analytic expressions}

Now we present analytic expressions for the single- and double-bubble spectrum.
In order to make the final expressions as simple as possible, 
we shift the origin of time to eliminate the linear term in Eq.~(\ref{eq:Gamma}):
\begin{align}
\Gamma(t)
&= \Gamma_*'' e^{- \gamma^2 t''^2},
\;\;\;\;
\Gamma_*''
= 
\Gamma_* e^{\frac{\beta^2}{4\gamma^2}}
= 
\beta'^4 e^{\frac{\beta'^2}{4\gamma^2}},
\;\;\;\;
t''
= 
t - \frac{\beta}{2\gamma^2}.
\label{eq:Gammaprpr}
\end{align}
In the last two equations, we have presented relations to the other parameterizations.
See Table~\ref{tbl:Param}.
The spectral shape calculated from this parameterization 
has a dependence on the dimensionless combination $\Gamma_*''/\gamma^4$,
which can be translated to a dependence on $\gamma/\beta'$ through the second equation
in Eq.~(\ref{eq:Gammaprpr}).

Now, following the basic strategy illustrated in the previous subsection,
we obtain (see Appendix~\ref{app:Derivation} for the details of the derivation)
\begin{align}
\Delta^{(s)}
&= 
\beta^2 v^6 k^3 \;
\Gamma_*''
\int_{-\infty}^\infty dt''_{\left< x,y \right>}
\int_{-\infty}^\infty dt''_{x,y}
\int_{|t''_{x,y}|}^\infty dr_v 
\nonumber \\
&\;\;\;\;\;\;\;\;\;\;\;\;\;\;\;\;\;\;
e^{-I(x,y)}
\left[
j_0(vkr_v){\mathcal S}_0
+ \frac{j_1(vkr_v)}{vkr_v}{\mathcal S}_1 
+ \frac{j_2(vkr_v)}{(vkr_v)^2}{\mathcal S}_2
\right]
\cos(kt''_{x,y}),
\label{eq:DeltaS}
\end{align}
for the single-bubble spectrum and 
\begin{align}
\Delta^{(d)}
&= 
\beta^2 v^9 k^3 \;
\Gamma_*''^2
\int_{-\infty}^\infty dt''_{\left< x,y \right>}
\int_{-\infty}^\infty dt''_{x,y}
\int_{|t''_{x,y}|}^\infty dr_v \;
e^{-I(x,y)}
\left[
\frac{j_2(vkr_v)}{(vkr_v)^2}{\mathcal D}_2
\right]
\cos(kt''_{x,y}),
\label{eq:DeltaD}
\end{align}
for the double-bubble spectrum.
Here ${\mathcal S}_0$, ${\mathcal S}_1$, ${\mathcal S}_2$ and ${\mathcal D}_2$ are 
defined in Eqs.~(\ref{eq:AppS}) and (\ref{eq:AppD}),
and $j_0$, $j_1$ and $j_2$ are the spherical Bessel functions given by Eq.~(\ref{eq:Appj}). 
Also, $I$ is given by Eq.~(\ref{eq:AppI}).

We numerically checked that $\Delta^{(s)}(k)=\mathcal{O}(10^{0-1}) \times\Delta^{(d)}(k)$ and $k_{\text{peak}}^{(s)}\simeq k_{\text{peak}}^{(d)}$ for $\gamma/\beta'=10^{-1.0},\,10^{-0.5},\,10^{0.0},\,10^{0.5},\,10^{0.75}$ with $v=0.3,\,1.0$. It is a bit counter-intuitive that two contributions are in a similar order of magnitude. The spectrum depends on a correlator $\Pi(t_x, t_y, k)$ which is a Fourier transform of $\left< T(x)T(y) \right>$, and $\left< T(x)T(y) \right>$ is an integration of a probability times $T(x)T(y)$ for each configuration. A survival probability of the probability part is common, and in both ``single-bubble'' and ``double-bubble'' cases, the wall thickness $l_B$s are totally cancelled out. There are two factors making a difference. The first one is an extra factor $\Gamma'_* r^3 \Delta t = \beta'^4 r^3 \Delta t \lesssim 1$ of a double-bubble case. $r$ is a representative radius of bubbles in the configuration, and $\Delta t$ is a time scale of consideration. The second one is an angular integration over nucleation points. Double-bubble nucleation points more destructively interfere with each other in the angular integration over nucleation points after the Fourier transform, thus it makes suppression. However, neither of the two suppressions is significant. They are expected to make $O(1)$ differences as numerical results show.

\section{Numerical results}
\label{sec:Numerical}
\setcounter{equation}{0}

In this section we show the results for numerical evaluation of the spectrum 
(\ref{eq:DeltaS}) and (\ref{eq:DeltaD}).
We have used a multi-dimensional integration algorithm VEGAS in the CUBA library~\cite{Hahn:2004fe}.
In presenting the results, we define the spectrum normalized by 
its peak wavenumber $k_{\rm peak}$ and amplitude $\Delta_{\rm peak}$:
\begin{align}
\tilde{\Delta} (\tilde{k})
&\equiv
\frac{\Delta(k)}{\Delta_{\rm peak}},
\;\;\;\;
\tilde{k}
\equiv
\frac{k}{k_{\rm peak}}.
\label{eq:Deltatilde}
\end{align}
By definition, $\tilde{\Delta}$ has its peak at $\tilde{k} = 1$.
This quantity makes it easier to compare the spectral shape.
Also, since we would like to know the deviation of the spectral shape from the one with $\gamma = 0$,
we define the ratio $R(\tilde{k})$ between the normalized spectra $\tilde{\Delta}$ 
with $\gamma = 0$ and $\gamma \neq 0$:
\begin{align}
R(\tilde{k})
&\equiv
\frac{\tilde{\Delta}(\tilde{k})}{\tilde{\Delta}_{\gamma = 0}(\tilde{k})}.
\label{eq:R}
\end{align}
Here $\tilde{\Delta}_{\gamma = 0}$ denotes the spectrum $\tilde{\Delta}$ with $\gamma = 0$.

\begin{figure}[t!]
\begin{center}
\includegraphics[height=0.28\columnwidth]{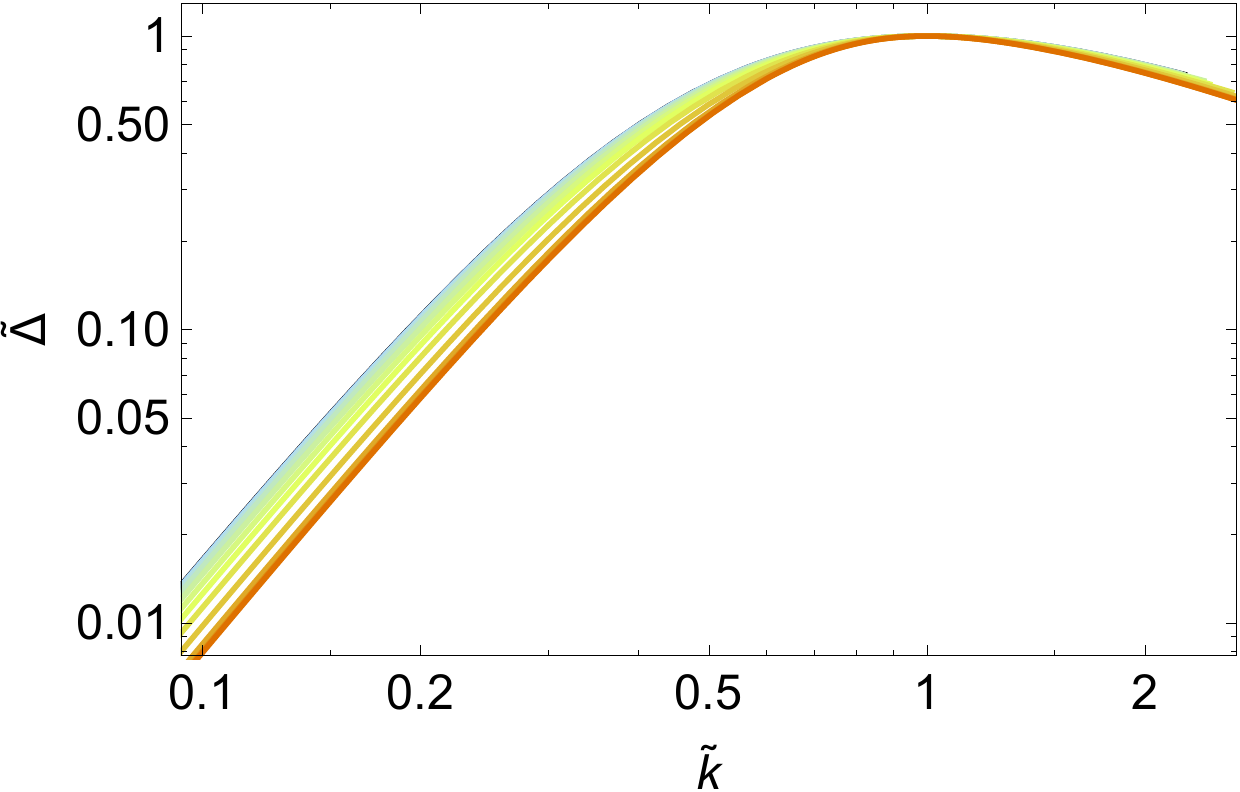} 
\quad
\includegraphics[height=0.28\columnwidth]{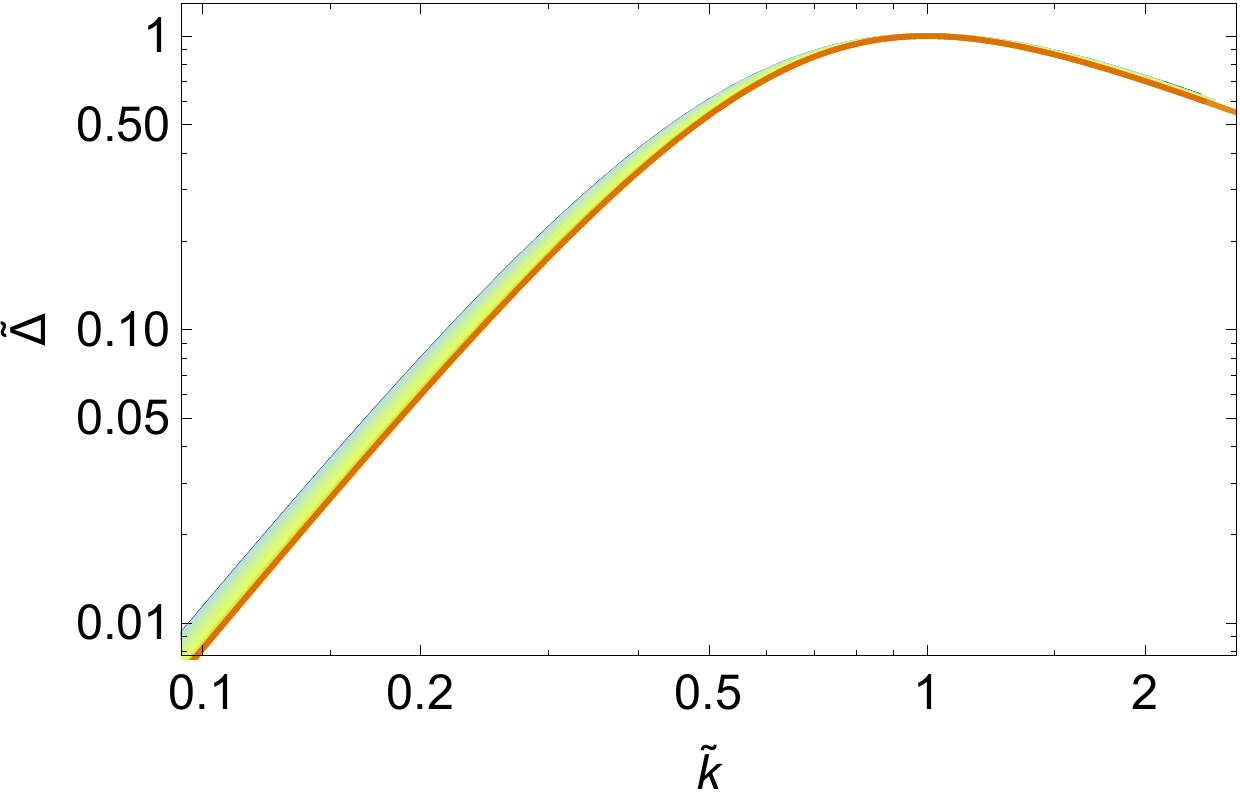} 
\includegraphics[height=0.28\columnwidth]{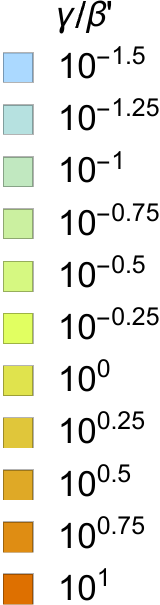} 
\caption{\small
(Left) Spectral shape for $v = 1$.
(Right) Spectral shape for $v = 0.3$.
}
\label{fig:SpectralShape}
\end{center}
\end{figure}

Now we present the results.
First, we show the normalized total spectrum $\tilde{\Delta}$ for various values of $\gamma/\beta'$
in Fig.~\ref{fig:SpectralShape}.
The wall velocity is taken to be $v = 1$ (left) and $v = 0.3$ (right), respectively.
Observed features are
\begin{itemize}
\item
The spectral shape starts to deviate from the one with $\gamma = 0$ 
for $\gamma/\beta' \sim {\mathcal O}(0.1)$.
\item
The spectral shape approaches to an asymptotic form for $\gamma/\beta' \gg {\mathcal O}(0.1)$.
\end{itemize}
The latter behavior is because the setup reduces to the $\delta$-function nucleation rate 
$\Gamma(t) \propto \delta(t)$: see Appendix~\ref{app:AppDelta} for details.

\begin{figure}[t!]
\begin{center}
\includegraphics[height=0.28\columnwidth]{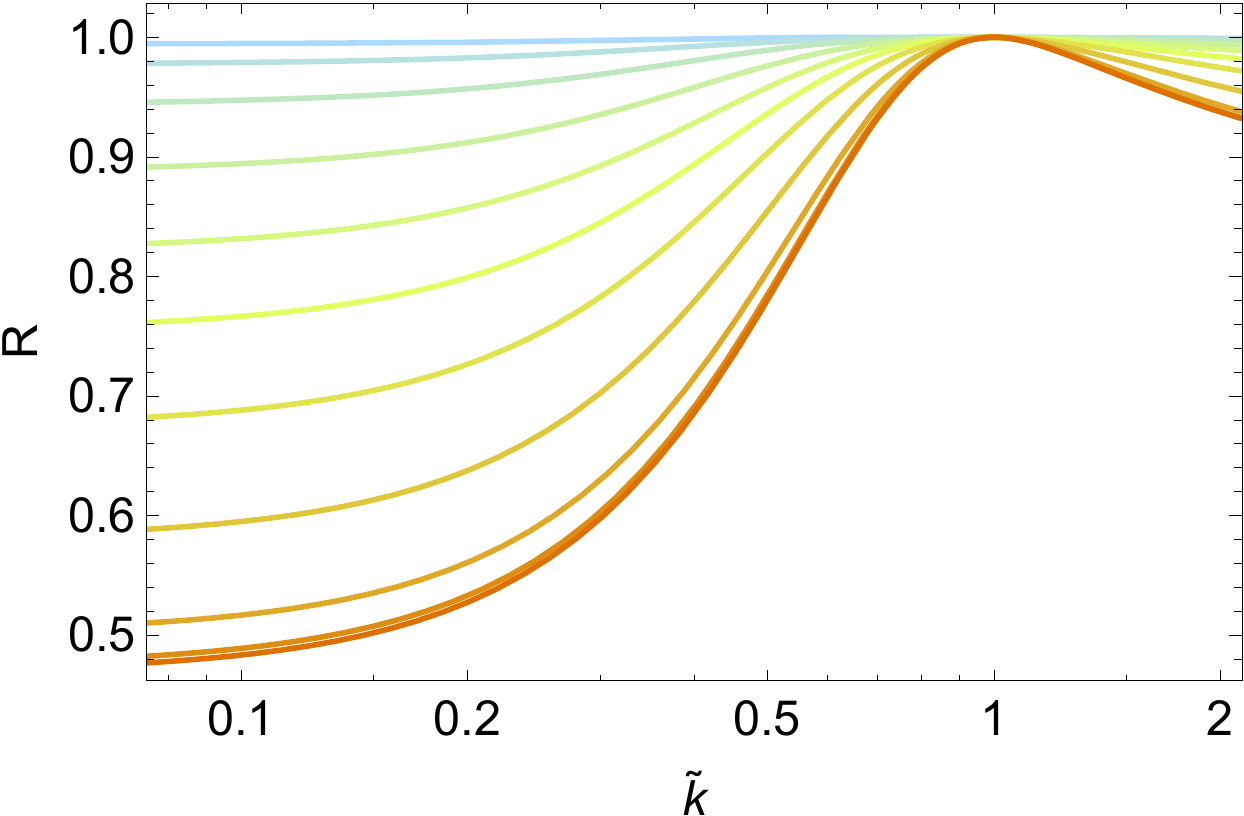} 
\quad
\includegraphics[height=0.28\columnwidth]{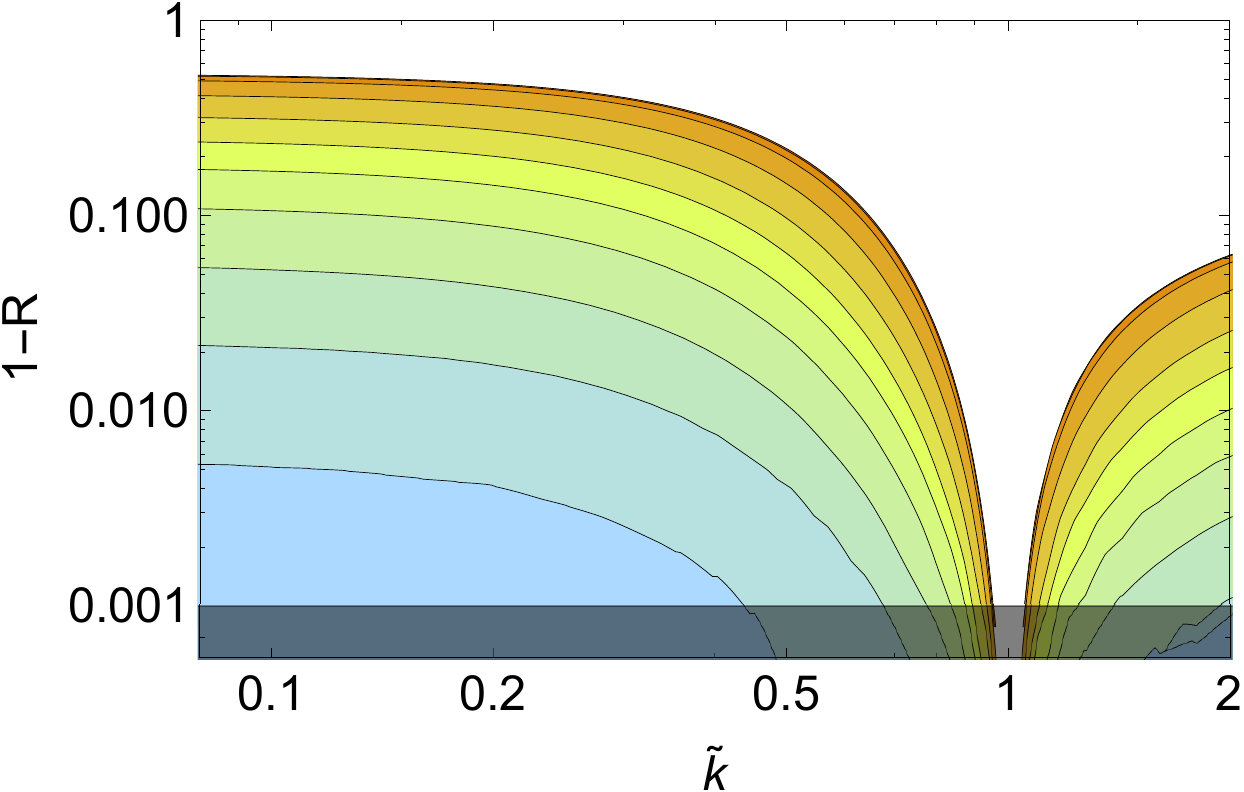} 
\includegraphics[height=0.28\columnwidth]{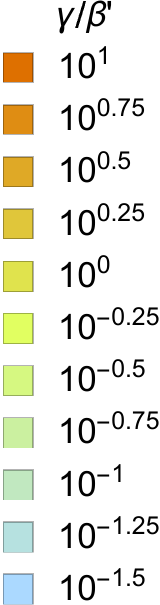} 
\caption{\small
(Left) 
Ratio $R$ between the spectra with $\gamma = 0$ and $\gamma \neq 0$ defined in Eq.~(\ref{eq:R}).
This figure shows $v = 1$ case for various values of $\gamma/\beta'$.
(Right) Log plot of the left panel.
}
\label{fig:SpectralShape_v=1}
\end{center}
\begin{center}
\includegraphics[height=0.28\columnwidth]{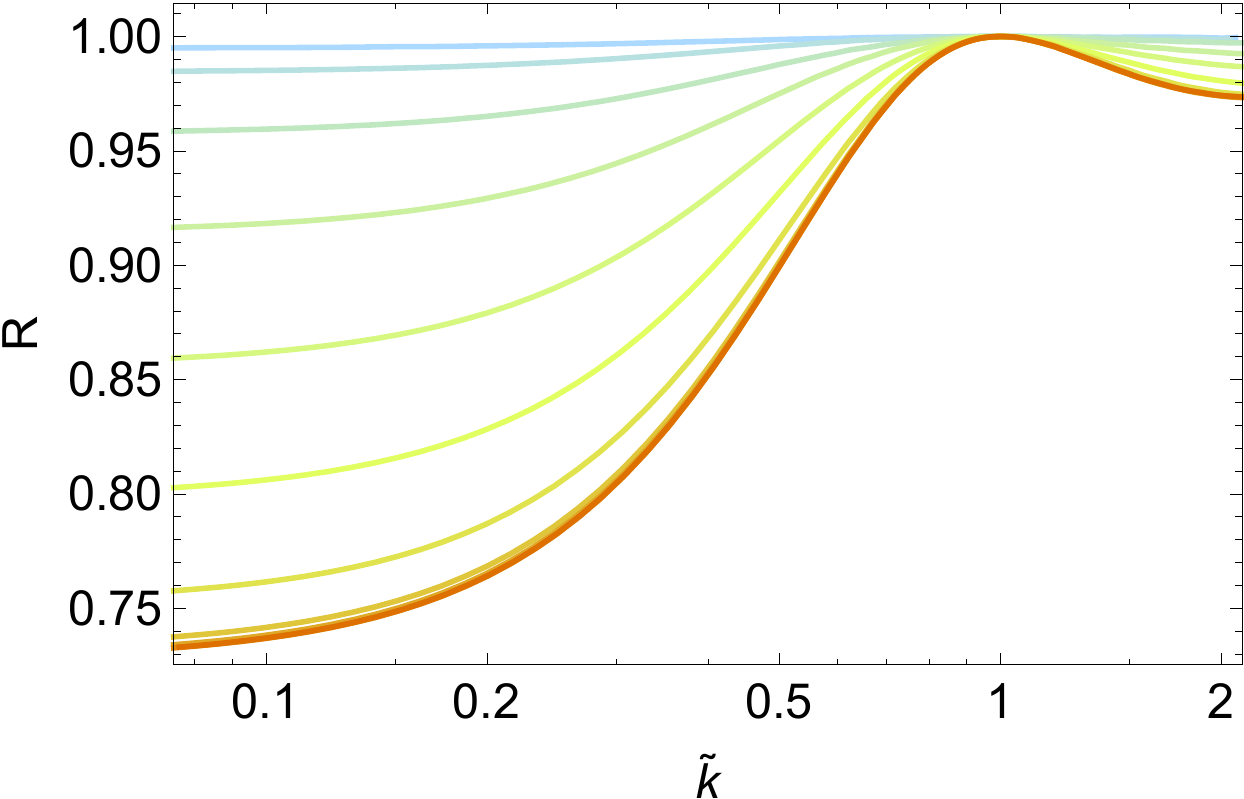} 
\quad
\includegraphics[height=0.28\columnwidth]{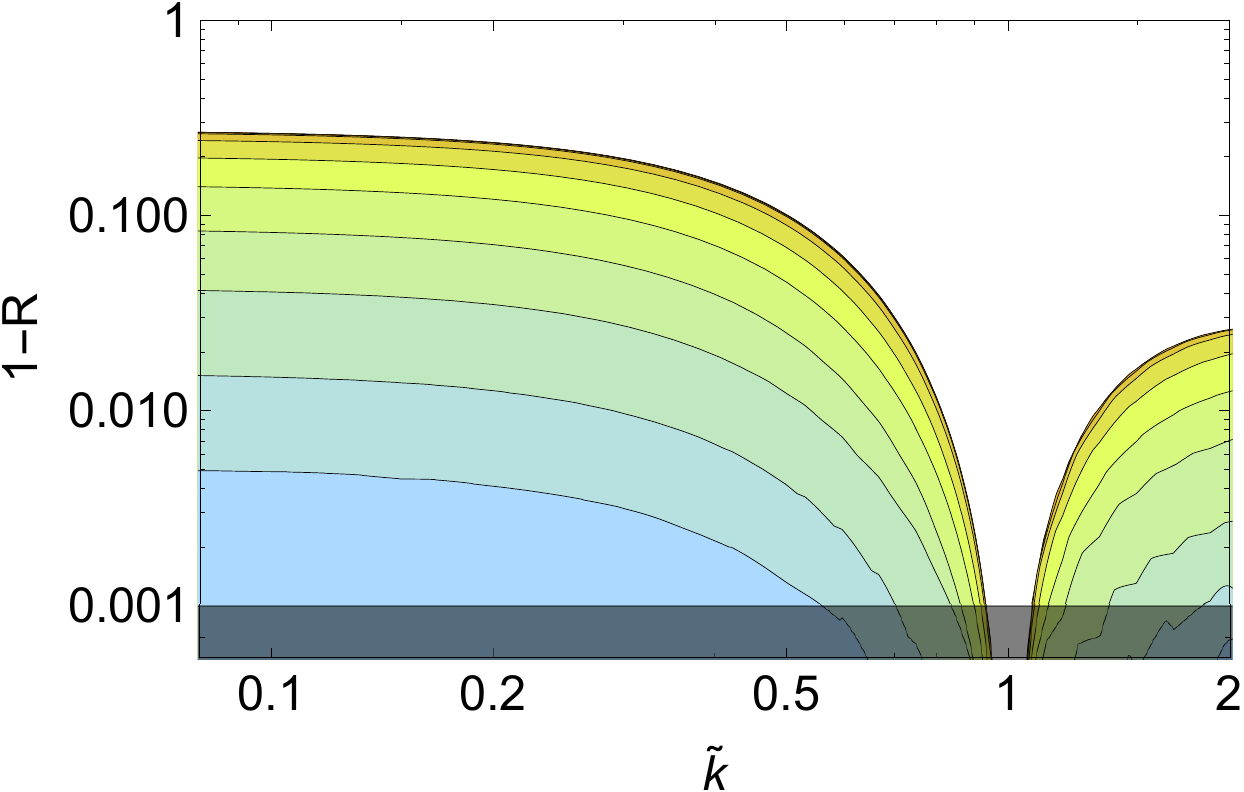} 
\includegraphics[height=0.28\columnwidth]{./figs/HelloWorld.pdf} 
\caption{\small The same as Fig.~\ref{fig:SpectralShape_v=1} except that $v = 0.3$.}
\label{fig:SpectralShape_v=0.3}
\end{center}
\end{figure}

Next, we plot the ratio $R$ for $v = 1$ and $v = 0.3$
in Figs.~\ref{fig:SpectralShape_v=1} and \ref{fig:SpectralShape_v=0.3}, respectively.
In these figures, the left panels show linear plots for the ratio $R$,
while the right panels are logarithmic plots for the deviation $1 - R$.
The black shaded regions on the right panels
represent the $0.1\%$ error coming from the cutoff for the Monte-Carlo integration.
Observed features are
\begin{itemize}
\item
For fixed $\gamma/\beta'$, the deviation in the spectral shape ({\it i.e.} $1 - R$) 
increases as $k$ deviates from $k_{\rm peak}$.
\item
The ratio $R$ approaches to constant in the small $k$ limit.
On the other hand, we have not confirmed such behavior in the large $k$ limit 
due to numerical difficulties,
though such a tendency is observed for example in Fig.~\ref{fig:SpectralShape_v=0.3} 
for large $\gamma/\beta'$. 
\item
For small $\gamma/\beta'$, the values of $R$ in $v=1$ and $v=0.3$ cases show similar deviation from unity. 
For larger $\gamma/\beta'$, however, the values of $R$ in $v=1$ case show a larger deviation from unity 
than $v=0.3$ case.
\end{itemize}

\begin{figure}[t!]
\begin{center}
\includegraphics[width=0.6\columnwidth]{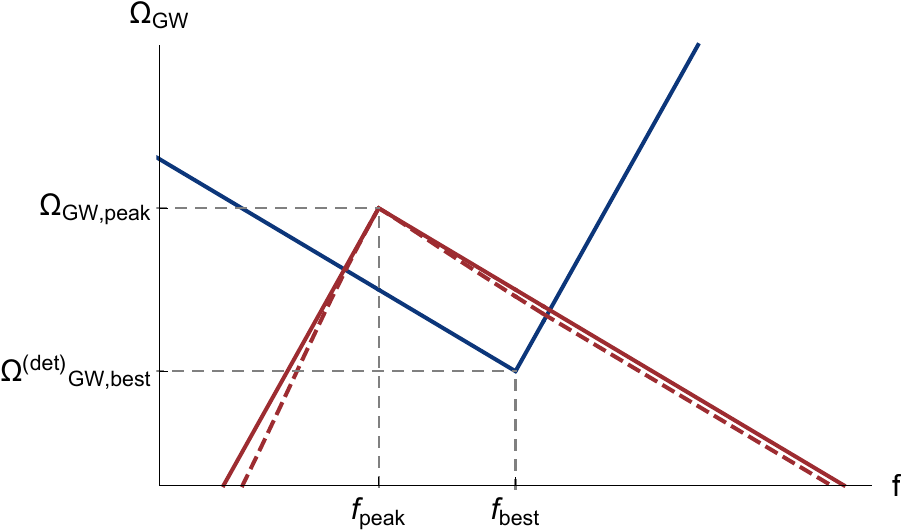} 
\caption{\small
Setup in Eqs.~(\ref{eq:Detector})--(\ref{eq:Distinguish}).
The blue line denotes the detector sensitivity curve (\ref{eq:Detector}),
while the red lines correspond to the signal $\Omega_{{\rm GW}, \gamma = 0}$ (solid)
and $\Omega_{\rm GW}$ with $\gamma / \beta' = 5$ (dashed).
The bubble wall velocity is taken to be $v = 1$,
and we have extrapolated the normalized spectrum $R(f/f_{\rm peak})$ 
below $f/f_{\rm peak} < 0.1$ and $f/f_{\rm peak} > 4$
by assuming that it is constant for these frequencies. The extrapolation is expected to give conservative estimate for the deviation.
}
\label{fig:Setup}
\vskip 0.6in
\includegraphics[width=0.8\columnwidth]{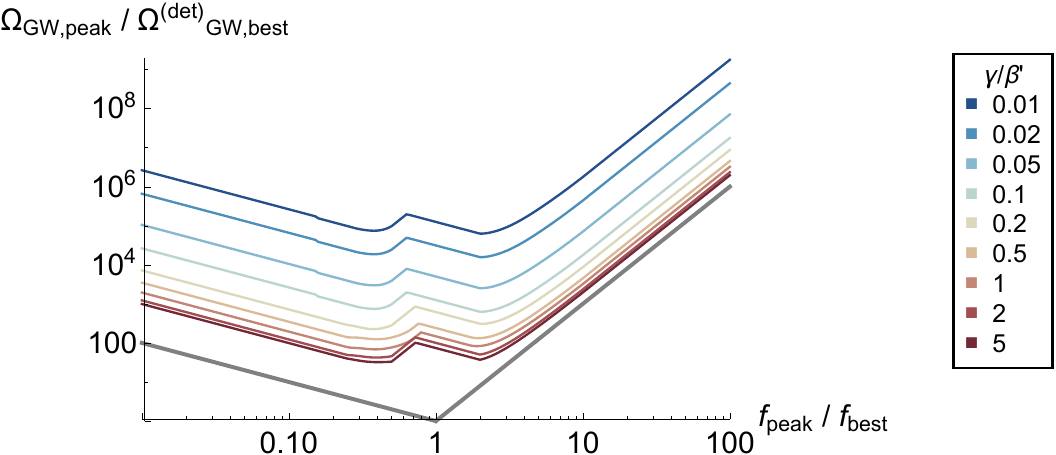} 
\caption{\small
Contour plot of $\gamma/\beta'$ above which 
the distinguishable condition (\ref{eq:Distinguish}) 
for the difference between the spectral shape with $\gamma \neq 0$ and $\gamma = 0$ is satisfied.
The detector sensitivity is given by Eq.~(\ref{eq:Detector}), 
while the signal shape is given by Eq.~(\ref{eq:SignalH}).
The wall velocity is taken to be $v = 1$ and 
$\gamma / \beta'$ is varied from $0.01$ (top) to $5$ (bottom).
The gray line corresponds to the boundary where the condition
${}^\exists f$ : $\Omega_{\rm GW}^{\rm (det)} (f) = \Omega_{{\rm GW}, \gamma = 0} (f)$
is satisfied, {\it i.e.}, detection of the spectrum with $\gamma = 0$.
}
\label{fig:Distinguish}
\end{center}
\end{figure}

Finally, let us briefly discuss the distinguishability of the spectral shape by future observations.
We assume the detector sensitivity to be
\begin{align}
\Omega_{\rm GW}^{\rm (det)} (f)
&= 
\Omega_{\rm GW, best}^{\rm (det)}
\times
\left\{
\begin{matrix}
\displaystyle \left( \frac{f}{f_{\rm best}} \right)^{-1} 
\;\;\;\; &(f < f_{\rm best}), \\
\displaystyle \left( \frac{f}{f_{\rm best}} \right)^3
\;\;\;\; &(f > f_{\rm best}),
\end{matrix}
\right. 
\label{eq:Detector}
\end{align}
while we approximate the signal by
\begin{align}
&\Omega_{\rm GW} (f)
= 
R
\left( \frac{f}{f_{\rm peak}} \right)
\times 
\Omega_{{\rm GW}, \gamma = 0} (f), 
\label{eq:SignalH}
\\
&\Omega_{{\rm GW}, \gamma = 0} (f)
= 
\Omega_{\rm GW, peak}
\times
\left\{
\begin{matrix}
\displaystyle \left( \frac{f}{f_{\rm peak}} \right)^3
\;\;\;\; &(f < f_{\rm peak}), \\
\displaystyle \left( \frac{f}{f_{\rm peak}} \right)^{-1}
\;\;\;\; &(f > f_{\rm peak}).
\end{matrix}
\right. 
\label{eq:Signal}
\end{align}
Here $f$ denotes the GW frequency,
and the high and low frequency behavior of $\Omega_{\rm GW}^{\rm (det)}$ 
models the shot noise and the radiation pressure noise, 
respectively~\cite{Maggiore:1900zz}.
Note that the argument of $R$ satisfies $f/f_{\rm peak} = k/k_{\rm peak}$. 
We illustrate the setup in Fig.~\ref{fig:Setup}.
In drawing the red-dashed line 
we have taken $v = 1$ and extrapolated the ratio $R(f/f_{\rm peak})$
for $f/f_{\rm peak} < 0.1$ and for $f/f_{\rm peak} > 4$
by assuming that it is constant for these frequencies.
Now let us define a condition for spectral shapes to be 
distinguished from each other by observations.
We call the spectral shapes ``distinguishable" if
\begin{align}
{}^\exists f
\;\;\;
:
\;\;\;
\Omega_{\rm GW}^{\rm (det)} (f)
<
\Delta \Omega_{\rm GW} (f)
\equiv
\Omega_{{\rm GW}, \gamma = 0} (f) - \Omega_{\rm GW} (f).
\label{eq:Distinguish}
\end{align}
This means that,
for fixed peak frequency and amplitude $(f_{\rm peak},\Omega_{\rm GW,peak})$,
a detector with $(f_{\rm best},\Omega_{\rm GW,best}^{\rm (det)})$ 
can distinguish two spectral shapes with $\gamma \neq 0$ and $\gamma = 0$.
This is expected to give a rough estimate for the sensitivity to the value of $\gamma$.

Fig.~\ref{fig:Distinguish} is a contour plot for $\gamma/\beta'$ 
above which the distinguishable condition is satisfied.
In this figure we have taken $v = 1$ and varied $\gamma / \beta'$ from $0.01$ to $5$.
In making this figure, 
we have extrapolated 
the values of $R$ 
for
$\gamma/\beta' < 0.1$
from
$\gamma/\beta' \geq 0.1$,
due to numerical difficulties arising for small $\gamma/\beta'$
(see Figs.~\ref{fig:SpectralShape_v=1}--\ref{fig:SpectralShape_v=0.3}).
The extrapolation procedure is as follows.
Regarding the Gaussian correction to the nucleation rate
as a perturbation to the standard nucleation rate $\Gamma \propto e^{\beta t}$,
one sees that the perturbation is controlled by $\gamma^2$.
Noting that the spectral shape is affected only by dimensionless quantities,
one expects that the deviation in the spectral shape from $\gamma = 0$ case is proportional to $\gamma^2/\beta'^2$
for small values of $\gamma/\beta'$.
We have confirmed this behavior at some fixed wavenumber,
and we present the details in Appendix~\ref{app:Asymptotic}.
From this observation, we have extrapolated the values of $R$ 
by using $1- R \propto \gamma^2/\beta'^2$ for small values of $\gamma/\beta'$.
In Fig.~\ref{fig:Distinguish}, 
it is shown that the spectral shape is distinguishable for
$\Omega_{\rm GW,peak}/\Omega_{\rm GW,best}^{\rm (det)} \gtrsim {\mathcal O}(100)$
for moderate values $\gamma / \beta' \sim {\mathcal O}(0.1)$.\footnote{
Actual sensitivity of GW detectors can be much better than the sensitivity curve.
This is because, for cross-correlation detectors, the signal-to-noise ratio 
improves with the observation period $T_{\rm obs}$:
$(S/N)^2 \sim T_{\rm obs} \int df \; \left[ \Omega_{\rm GW}(f)/\Omega_{\rm GW}^{\rm (det)}(f) \right]^2$
(see {\it e.g.} Ref.~\cite{Allen:1997ad}),
while the usual sensitivity curve does not take into account this improvement
(see also Ref.~\cite{Thrane:2013oya} on this point).
We do not go into such details and just use 
the setup (\ref{eq:Detector})--(\ref{eq:Distinguish}) for simplicity.
}

\section{Conclusions}
\label{sec:Conclusion}
\setcounter{equation}{0}

In this paper, we studied gravitational-wave (GW) production in the cosmic first-order phase transition.
As stressed in Introduction, 
the GW spectrum resulting from the bubble dynamics contains information on
\begin{itemize}
\item[(1)]
Spacetime distribution of bubbles ({\it i.e.}, nucleation rate of bubbles),
\item[(2)]
Energy-momentum tensor profile around a bubble wall,
\item[(3)]
Dynamics after bubble collisions.
\end{itemize}
In maximizing the amount of information on the unknown high-energy physics 
extracted from the spectrum, all these three aspects cannot be missed.
In this paper we focused on the effect of (1) on the spectrum.
For this purpose we used the method of relating the GW spectrum
with the two-point ensemble average of the energy-momentum tensor $\left< T(x)T(y) \right>$.
As pointed out in Ref.~\cite{Jinno:2016vai},
all the contributions to this two-point correlator reduce to two classes
under the thin-wall approximation (which refers to (2)).
With the envelope approximation (which refers to (3)),
we wrote down analytic expressions for the spectrum 
with the nucleation rate $\Gamma(t) \propto e^{\beta t - \gamma^2 t^2}$,
and investigated the effect of the Gaussian correction $\gamma$ on the spectrum.

As a result, we found that the spectral shape differs from the one without the Gaussian correction
by ${\mathcal O}(10)\%$ for moderate values of $\gamma$ ($\gamma/\beta \sim {\mathcal O}(0.1))$,
see Figs.~\ref{fig:SpectralShape}--\ref{fig:SpectralShape_v=0.3}.\footnote{
In these figures we used $\beta'$ instead of $\beta$;
see Sec.~\ref{sec:Formalism} for the definition.
Also, for $\gamma/\beta'<{\mathcal O}(0.1)$, $\gamma/\beta'\simeq\gamma/\beta$ holds;
see Fig.~\ref{fig:GammaGammapr}.
}
These values for the Gaussian correction are typically expected in various models
as detailed in Appendix~\ref{app:Model}.
Therefore, as shown in Figs.~\ref{fig:Setup}--\ref{fig:Distinguish}, 
we have chances for extracting the information on the nucleation rate from the spectrum
if the signal is well above the sensitivity of the GW detectors.

Model separation by the information encoded in the GW spectral shape is one of the most important tasks in this field.
In this paper, we have shown that we can extract information about
the Gaussian correction to the nucleation rate in a simplified setup.
Though much remains to be done for model separation in a realistic setup,
this direction will be worth investigating further in the future.

\section*{Acknowledgments}

The work of R.J. and S.L. was supported by IBS under the project code, IBS-R018-D1.
The work of M.T. was supported by JSPS Research Fellowships for Young Scientists.

\appendix

\section{Typical values of $\gamma/\beta'$}
\label{app:Model}
\setcounter{equation}{0}

In this appendix, we discuss typical values of the parameter $\gamma/\beta'$.
After discussing some general aspects,
we estimate $\gamma/\beta'$ for some motivated models.
We will see that the typical value is around $\mathcal{O}(0.1)$~\footnote{
As mentioned in Sec.~\ref{subsec:AA}, 
much larger $\gamma/\beta'$ can be realized 
in some models~\cite{Espinosa:2008kw,Huber:2015znp,Leitao:2015fmj,Megevand:2016lpr,Kobakhidze:2017mru,Cai:2017tmh}.
}
For notational simplicity, we write $\gamma/\beta'$ as $\gamma/\beta$ in the following.

First, let us fix notations.
In finite temperature field theory, the bubble nucleation rate per unit four volume
is given by~\cite{Linde:1977mm,Linde:1981zj}
\begin{align}
         \Gamma(T)=b(T)T^4e^{-S_3(T)/T},
\end{align}
where $S_3$ denotes the three dimensional bounce action
and $b(T)\sim \mathcal{O}(1)$ denotes a contribution from the prefactor.
Since the dependence of $\Gamma(T)$ on $b(T)$ is generically very weak 
compared to other dependences, we set $b(T)=1$.
We rewrite the nucleation rate as
\begin{align}
         \Gamma(T)&=M^4e^{-S(T)}, \\
         S(T)&=\frac{S_3(T)}{T}-4\log(T/M),
\end{align}
where we have introduced some typical mass scale $M$ of a given model.
At a given temperature $T$, we can relate $S(T)$ with $\beta$ and $\gamma$,
which are expansion parameters of the nucleation rate in terms of time 
($\Gamma(t) \propto e^{\beta t-\gamma^2t^2}$).
By noticing $dT/dt = -HT$, we have
\begin{align}
\frac{\beta}{H}
&=
T\frac{dS(T)}{dT},
\label{eq:betaH} \\
\frac{\gamma^2}{H^2}
&=
\frac{1}{2}
\left[
\frac{\beta}{H}\left( 1 + \frac{T}{H}\frac{dH}{dT} \right) + T^2\frac{d^2S(T)}{dT^2}
\right].
\label{eq:gammaH}
\end{align}
We use parameters $\beta/H$ and $\gamma/\beta$ instead of $\beta$ and $\gamma$
since they can be estimated only by the thermal field theory,
that is, they can be calculated only from the bounce action $S(T)$ 
without the Einstein equation, as we see from Eqs.~(\ref{eq:betaH}) and (\ref{eq:gammaH}).
As mentioned in Sec.~\ref{sec:nucr}, the phase transition mainly occurs when
$\Gamma(T)\simeq \beta^4$. We define the transition temperature $T_*$
by the following condition
\begin{align}
         \Gamma(T_*)=\beta(T_*)^4.
\end{align}
We can rewrite this condition as follows:
\begin{align}
         S(T_*)=4\log\left[\frac{M}{H(T_*)}\right]+4\log\left[\frac{H(T_*)}{\beta(T_*)}\right].
\label{eq:s}
\end{align}
For later convenience, we introduce $S_C(T)$ (with $C$ denoting ``critical") as
\begin{align}
         S_C(T)\equiv 4\log\left[\frac{M}{H(T)}\right]+4\log\left[\frac{H(T)}{\beta(T)}\right].
\label{eq:sc}
\end{align}
Now the transition condition, which determines $T_*$, is simply given by $S(T)=S_C(T)$.

Next, let us consider typical values of $\gamma/\beta$, 
which determine the spectral shape of GWs.
Let us approximate
the transition rate around the transition temperature by the following form
\begin{align}
         S=A(T+B)^n,
\end{align}
with some constants $A$ and $B$ and $\mathcal{O}(1)$ constant $n$.
When the derivative of the Hubble parameter in Eq.~(\ref{eq:gammaH}) is negligible,\footnote{
This condition is satisfied for example when $\beta/H \gg 100$ 
(because $1/S$ dominates in Eq.~(\ref{eq:gammabeta}))
or when the vacuum energy dominates the radiation energy
(because $1$ dominates in Eq.~(\ref{eq:gammaH})).
The former occurs in Appendix~\ref{App:Singlet} and \ref{App:Flaton},
while the latter occurs in Appendix~\ref{App:Clacon} and \ref{App:Flaton}.
}
$\gamma/\beta$ at a given temperature $T$ is given by
\begin{align}
\frac{\gamma}{\beta}
&=
\sqrt{\frac{H}{2\beta}+\frac{n-1}{2n}\frac{1}{S(T)}}.
\label{eq:gammabeta}
\end{align}
Since at the time of transition $\beta/H \sim 10^{1-5}$ and $S(T) \sim 100$ hold 
in most cases of interest, 
we expect $\gamma/\beta \sim \mathcal{O}(0.1)$ typically.

Below, we estimate $\gamma/\beta$ for some motivated models.
We consider three models: 
\begin{itemize}
\item[(1)] 
Singlet extension of the standard model,
\item[(2)]
Classically conformal $B-L$ model,
\item[(3)] 
Supersymmetric flaton model.
\end{itemize}
We will see that $\gamma/\beta \simeq 0.08$, 
$\gamma/\beta \simeq 0.1$ -- $0.2$ and $\gamma/\beta \lesssim 0.04$ 
for the three cases, respectively.
Thus, if we have $\mathcal{O}(0.1)$ sensitivity on $\gamma/\beta$,
we can distinguish these models in principle.

\subsection{Singlet extension of the standard model}
\label{App:Singlet}

First let us consider singlet extension of the standard model.
Gravitational wave production in phase transitions in this type of model 
has been extensively studied in the literature.
Here we consider a simplified version of this model.
The potential is
\begin{align}
V_0
&=
- \frac{\mu^2}{2}h^2 
+ \frac{\lambda_H}{4}h^4
+ \sum_{i=1}^{N_s}
\left( 
\frac{m_s^2}{2}s_i^2 + \frac{\lambda_s^2}{2}h^2s^2_i
\right),
\end{align}
where $h$ represents the Higgs field, $s_i$ are real scalar fields and
$N_s$ denotes the number of singlets.
We take universal mass and coupling $m_s$ and $\lambda_s$ for these singlets.
For the effective potential, we take only the one-loop part from singlet loops 
by assuming that these contributions dominate over those from the standard model. 
We have
\begin{align}
         V_{\rm 1-loop}&=V_{\rm CW}+V_{\rm th},
\end{align}
where
\begin{align}
V_{\rm CW}(h)
&=
\frac{N_S(m_s^2+\lambda_s^2h^2)^2}{64\pi^2}\log
\left(\frac{m_s^2+\lambda_s^2h^2}{\Lambda^2}\right), \\
V_{\rm th}(h,T)
&=
N_SV_{\rm th}^B(x,T), 
\;\;\;\; x \equiv \sqrt{m_s^2+\lambda_s^2h^2}/T,\\
V_{\rm th}^{B/F}(x,T)
&=
\pm\frac{T^4}{2\pi^2}\int_0^\infty dz\;z^2\log\left[1\mp e^{-\sqrt{z^2+x^2}}\right],
\end{align}
where $V_{\rm CW}$ is the Coleman-Weinberg potential, 
$V_{\rm th}^{B/F}$ denotes thermal one-loop potential~\cite{Dolan:1973qd},
and $\Lambda$ denotes the renormalization scale.
Assuming $m^2_s\gtrsim \lambda_s^2T^2/6$, 
we neglect so-called daisy correction~\cite{Arnold:1992rz}.
We fix the vacuum expectation value and mass of $h$ to those of the standard-model Higgs boson 
$v_{\rm EW}$ and $m_H$.
For simplicity, we set $m_s^2/\lambda_s^2 = v^2_{\rm EW}/6$ 
and $\Lambda^2 = m_s^2+\lambda_s^2v_{\rm EW}^2$.
Then, we have the following conditions:
\begin{align}
\frac{\mu^2}{v_{\rm EW}^2}
&=
\frac{1}{2}\frac{m_H^2}{v_{\rm EW}^2} - \frac{11}{192\pi^2}N_s\lambda_s^4, \\
\lambda_H
&=
\frac{1}{2}\frac{m_H^2}{v_{\rm EW}^2} - \frac{3}{32\pi^2}N_s\lambda_s^4, 
\end{align}

\begin{figure}[t!]
\begin{center}
\includegraphics[width=0.7\columnwidth]{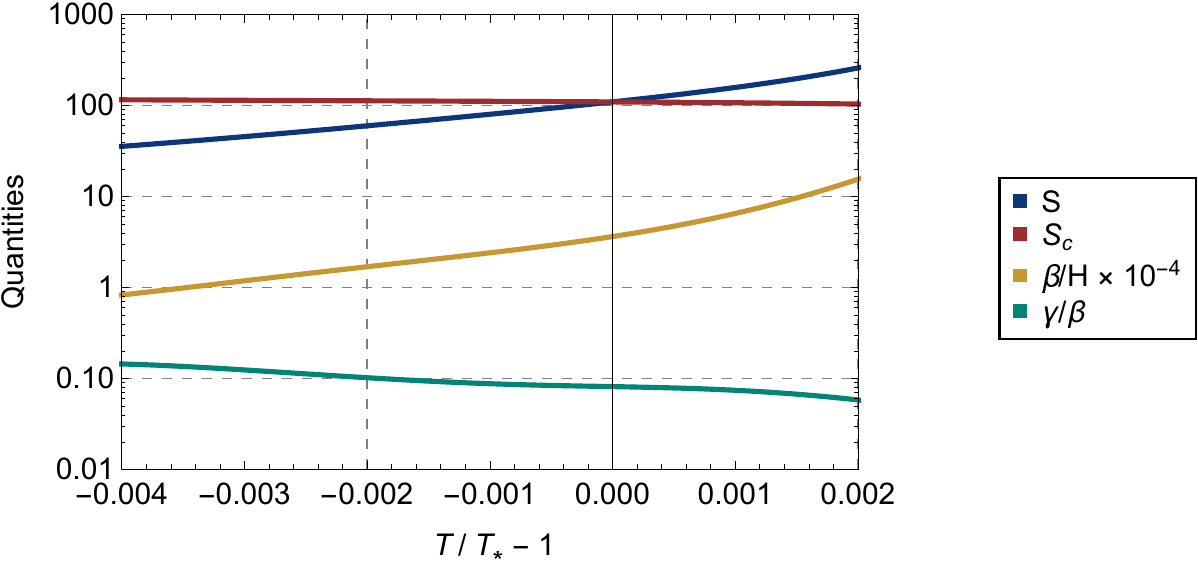} 
\caption{
Temperature dependence of $S$, $S_C$, $\beta/H$ and $\gamma/\beta$
for the singlet extension model in Appendix~\ref{App:Singlet}.
}
\label{fig:Singlet}
\end{center}
\end{figure}

As a benchmark point, we take $N_s=10$, $\lambda_s=0.8$.
Fig.~\ref{fig:Singlet} shows the temperature dependence of
$S$, $S_C$, $\beta/H$ and $\gamma/\beta$.
Here we have taken the typical mass scale to be $M = v_{\rm EW}$.
The transition temperature $T_*$ is determined from $S(T_*)=S_C(T_*)$ as $T_* \simeq 0.6v_{\rm EW}$. 
At the transition temperature we have $\beta/H \simeq 4 \times 10^4$ and $\gamma/\beta \simeq 0.08$.

\subsection{Classically conformal $B-L$ model}
\label{App:Clacon}

Next let us consider the classically conformal $B-L$ model~\cite{Iso:2009ss,Iso:2009nw}.
Gravitational wave production in such a scenario is considered in Refs.~\cite{Jinno:2016knw,Iso:2017uuu}.\footnote{
Recently the gauge dependence of this type of models has been discussed in Ref.~\cite{Chiang:2017zbz}.
}
In this model, we impose so-called ``classical conformal invariance" 
based on the argument in Ref.~\cite{Bardeen:1995kv},
and also add U$(1)_{B-L}$ gauge symmetry.
A complex scalar field $\Phi$ is introduced in order to break this U$(1)_{B-L}$
by its vacuum expectation value (VEV) and to induce the masses of the right handed neutrinos.
The tree level scalar potential is given by
\begin{align}
         V=\lambda_H|H|^4+\lambda |\Phi|^4-\lambda'|\Phi|^2|H|^2,
\end{align}
where only quartic couplings appear due to the assumption of the classical conformal invariance.
The VEV of the new field $M_{\Phi} \equiv \sqrt{2} \langle \Phi \rangle$ induces
a negative mass term for the Higgs field and the electroweak scale can be realized. 
Below, we discuss a thermal phase transition of the $\Phi$ field.

For simplicity, we take $M_\Phi=10$~TeV $(\gg v_{\rm EW})$
and assume small right handed Yukawa couplings compered to the $B-L$ gauge coupling $g_{B-L}$.
In such a case, the potential for $\Phi$ is mainly determined within
$\Phi$ and $B-L$ gauge boson sector.
The running of the quartic coupling $\lambda$ determines the potential for $\Phi$,
and we can realize a minimum at $M_\Phi=10$~TeV. 
Now, because of the assumption $M_\Phi=10$~TeV, 
the only free parameter reduces the $B-L$ gauge coupling strength $\alpha_{B-L}(M_\Phi)$ 
at scale $M_\Phi$. 

\begin{figure}[t!]
\begin{center}
\includegraphics[width=0.7\columnwidth]{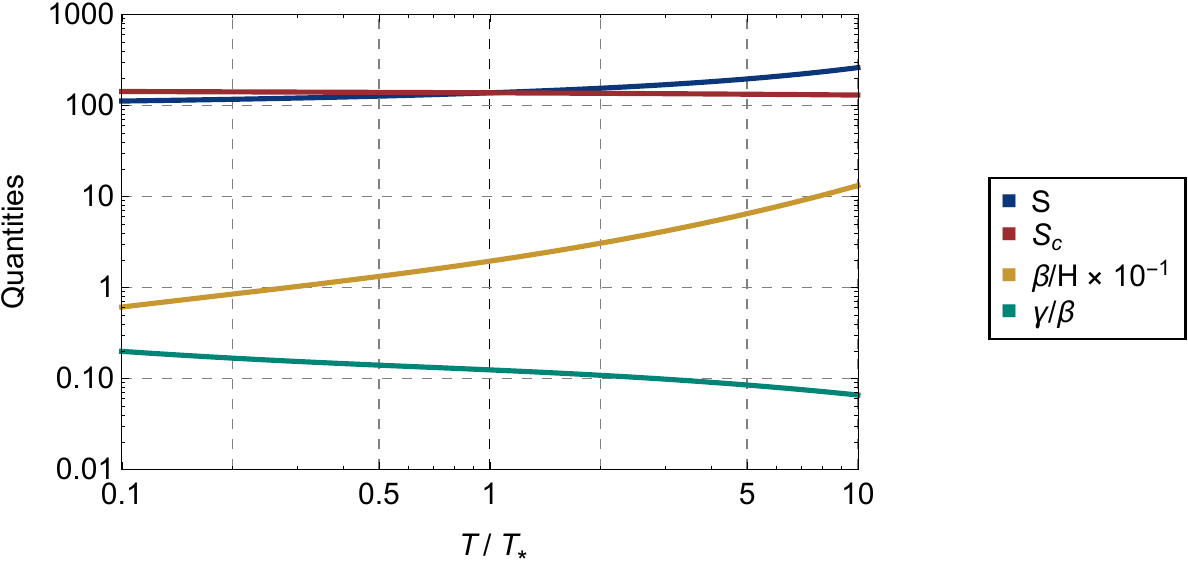} 
\caption{
Temperature dependence of $S$, $S_C$, $\beta/H$ and $\gamma/\beta$
with $\alpha_{B-L}(M_\Phi)=0.01$ for the classically conformal $B-L$ model.
}
\label{fig:Clacon1}
\end{center}
\vskip 0.2in
\begin{center}
\includegraphics[width=0.65\columnwidth]{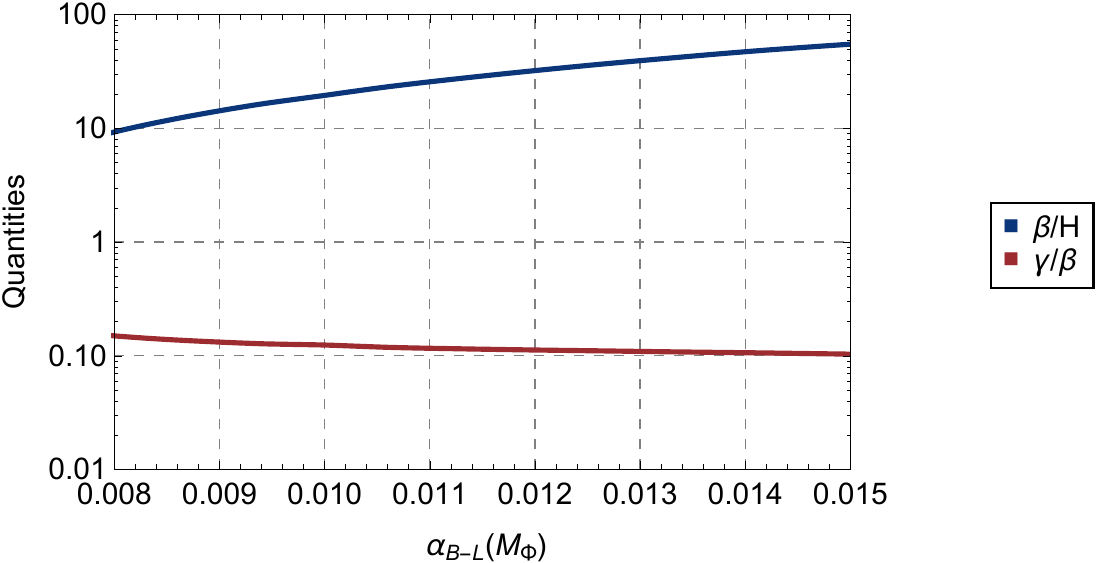} 
\caption{
$\alpha_{B-L}(M_\Phi)$ dependence of $\beta/H$ and $\gamma/H$ at  transition time
for the classically conformal $B-L$ model.
}
\label{fig:Clacon2}
\end{center}
\end{figure}

Figure~\ref{fig:Clacon1} shows the temperature dependence of
$S$, $S_C$, $\beta/H$ and $\gamma/\beta$ for $\alpha_{B-L}(M_\Phi)=0.01$.
Here we have used the effective potential in Ref.~\cite{Jinno:2016knw} and
taken the typical mass scale to be $M=M_\Phi$.
The transition temperature is obtained as $T_* = 5 \times 10^{-3}M_\Phi$,
and we have $\beta/H \simeq 20$ and $\gamma/\beta \simeq 0.12$ at this temperature.
We also consider $\alpha_{B-L}(M_\Phi)$ dependence of these parameters.
Fig.~\ref{fig:Clacon2} shows
$\alpha_{B-L}(M_\Phi)$ dependence of $\beta/H$ and $\gamma/H$ at the transition time.
We see that the parameter $\gamma/\beta$ varies in a range $\gamma/\beta \simeq 0.1-0.2$.\footnote{
If $\alpha_{B-L}(M_\Phi)$ is smaller than $\sim 0.008$, 
an ultra supercooling occurs and 
the QCD phase transition affects the dynamics~\cite{Iso:2017uuu}. 
We do not consider such a parameter region here for simplicity.
}

\subsection{Supersymmetric flaton model}
\label{App:Flaton}

Lastly we consider phase transitions after thermal inflation~\cite{Lyth:1995hj,Lyth:1995ka} 
in supersymmetric flaton models.
Gravitational wave production in such a scenario is studied in Ref.~\cite{Easther:2008sx}.
As in Ref.~\cite{Easther:2008sx}, we characterize the flaton potential by two parameters:
\begin{align}
         V(\phi)=V_{\rm TI}-\frac{1}{2}m_\phi^2\phi^2+\cdots,
\end{align}
where $m_\phi$ and $V_{\rm TI}$ denote the negative mass term and the potential energy around the origin,
respectively.
We assume that the potential shape is well approximated by this quadratic form for $\phi \lesssim m_\phi$. 
In addition, we assume that 
some higher-order terms stabilize the potential at $\phi \gg m_\phi$
and that we have $V(\langle \phi\rangle) = 0$.
In order to have a thermal potential, 
we add vector-like superfields $Q$ and $\bar{Q}$ with a superpotential 
\begin{align}
        W\supset \lambda \Phi \bar{Q}Q.
\end{align}
As a benchmark model,
we take $5$ and $\bar{5}$ representations of SU$(5)$ for $Q$ and $\bar{Q}$, respectively.
In this case, the one-loop thermal potential is given by
\begin{align}
&\;\;\;\;\;\;\;\;
V_{\rm th}
=
20V_{\rm th}^B(x,T) 
+ 20V_{\rm th}^{F}(y,T), \\
&x
\equiv 
\sqrt{m_\phi^2+\lambda^2\phi^2/2}/T, 
\;\;\;\;
y
\equiv 
\sqrt{\lambda^2\phi^2/2}/T,
\end{align}
where we have taken the mass of the scalar components of $Q$ and $\bar{Q}$ to be $m_\phi$, 
and neglected thermal masses for them.

\begin{figure}[t!]
\begin{center}
\includegraphics[width=0.7\columnwidth]{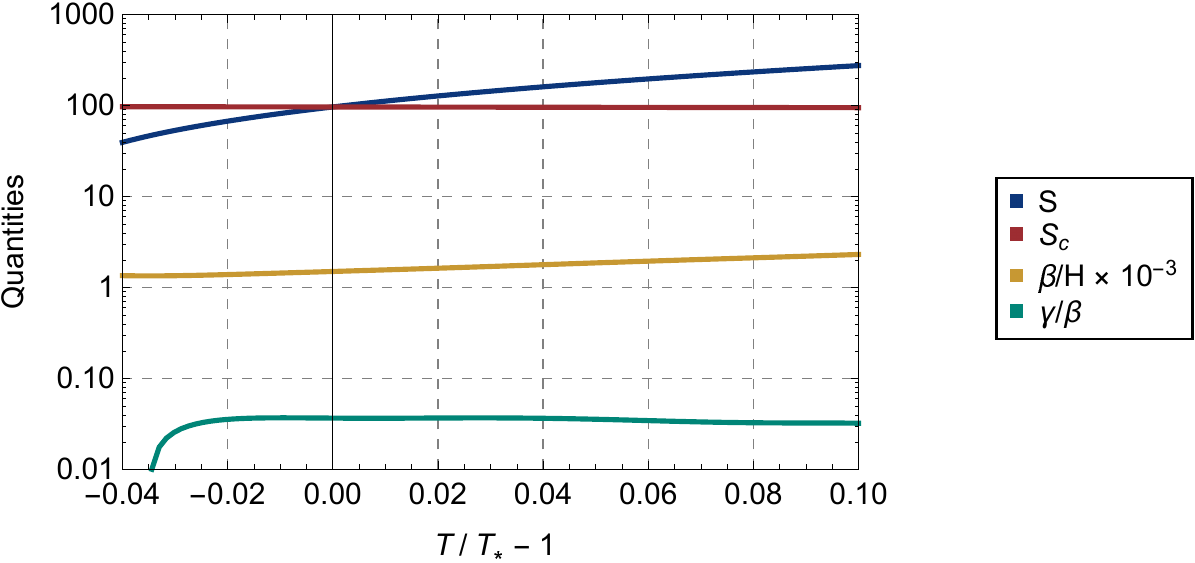} 
\caption{
Temperature dependence of $S$, $S_C$, $\beta/H$ and $\gamma/\beta$
with $\lambda = 0.8$, $V_{\rm TI}/m_\phi^4 = 10^4$ and $m_\phi = 1$ TeV
for the supersymmetric flaton model.
}
\label{fig:Flaton1}
\end{center}
\vskip 0.5in
\begin{center}
\includegraphics[width=0.55\columnwidth]{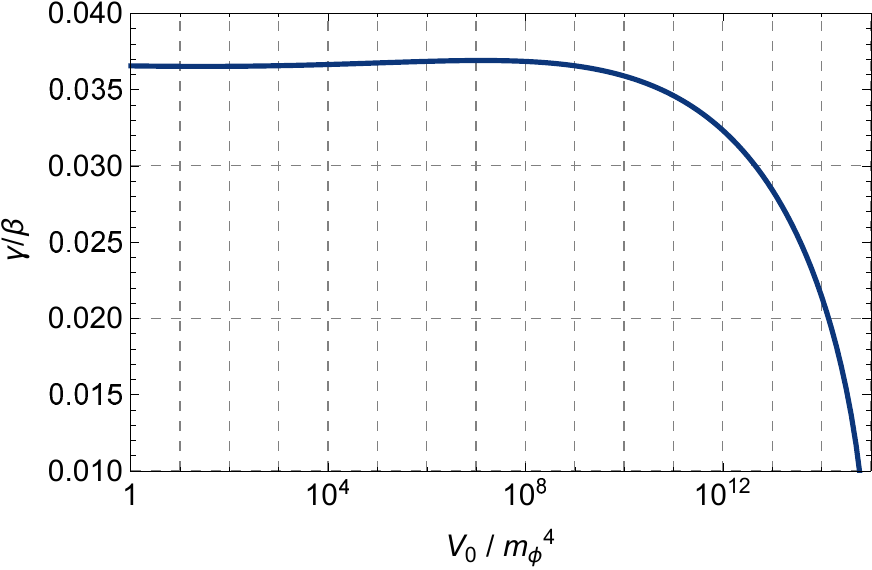} 
\caption{
$V_{\rm TI}/m_\phi^4$ dependence of $\gamma/\beta$ at the transition time with $\lambda = 0.8$
and $m_\phi = 1$ TeV
for the supersymmetric flaton model.
}
\label{fig:Flaton2}
\end{center}
\end{figure}

Now we have three free parameters: 
the coupling $\lambda$, the mass scale $m_\phi$, and the size of the vacuum energy $V_{\rm TI}/m_\phi^4$.
Fig.~\ref{fig:Flaton1} shows the temperature dependence of
$S$, $S_C$, $\beta/H$ and $\gamma/\beta$
for $\lambda=0.8$, $m_\phi=1$~TeV and $V_{\rm TI}/m_\phi^4 = 10^4$.
Here we have taken the typical mass scale to be $M = m_\phi$.\footnote{
Note that the relevant scale for the transition is $m_\phi$ rather than $V_{\rm TI}$,
because the former determines the mass scale of the potential around the origin.
}
The transition temperature is given by $T_*\simeq 1.4m_\phi$,
and we have $\beta/H\simeq 1500$ and $\gamma/\beta\simeq 0.04$ at the transition time.
In this setup, the sign of $\gamma^2$ flips at a lower temperature.
Fig.~\ref{fig:Flaton2} shows the $V_{\rm TI}/m_\phi^4$ dependence of $\gamma/\beta$ at the transition time 
with $\lambda=0.8$ and $m_\phi=1$~TeV. 
We have $\beta/H \simeq 1000-2000$ for this region.
We see that $\gamma/\beta$ decreases as $V_{\rm TI}/m_\phi^4$ increases. 
This is because the transition temperature $T_*$ becomes smaller for larger $V_{\rm TI}/m_\phi^4$:
for larger $V_{\rm TI}/m_\phi^4$,
$S_C$ becomes smaller because the Hubble parameter in the first term of Eq.~(\ref{eq:sc}) increases,
and the transition temperature $T_*$ given by $S = S_C$ decreases.
As a result, the sign of $\gamma^2$ at the transition time flips around $V_{\rm TI}/m_\phi^4 \sim 10^{15}$.
For the parameter values shown in this plot, 
$\gamma/\beta$ varies within $\gamma/\beta \lesssim 0.04$.

\clearpage

\section{Derivation of the analytic expressions}
\label{app:Derivation}
\setcounter{equation}{0}

In this appendix we derive the analytic expressions (\ref{eq:DeltaS}) and (\ref{eq:DeltaD}).
We use the nucleation rate 
shifted from the original form (\ref{eq:Gamma}) so that the linear term vanishes:
\begin{align}
\Gamma(t)
&= \Gamma_*'' e^{- \gamma^2 t''^2}.
\label{eq:AppGamma}
\end{align}
In the following we take $\gamma = 1$ unit.
Also, in this appendix we simply write $\Gamma_*''$ and $t''$ as $\Gamma_*$ and $t$, respectively,
for notational simplicity.

As mentioned in Sec.~\ref{sec:Analytic},
what we need in order to obtain the GW spectrum is to calculate $\left< T(t_x,\vec{x}) T(t_y,\vec{y}) \right>$,
with $x$ and $y$ denoting arbitrary four-dimensional spacetime points.
In the thin-wall limit, there are two classes of contributions to $\left< T(t_x,\vec{x}) T(t_y,\vec{y}) \right>$,
the single- and double-bubble, 
and thus the resulting GW spectrum can be classified into the single- and double-bubble spectra.
In the derivation, we consider only those configurations with $r_v > |t_{x,y}|$
where $\vec{r} \equiv \vec{x} - \vec{y}$, $r \equiv |\vec{r}|$, 
$r_v \equiv r/v$ and $t_{x,y} \equiv t_x - t_y$
because only such configurations are relevant under the envelope approximation.
First, we introduce the ``false vacuum probability" $P(x,y)$, 
and then proceed to the calculation of the single- and double-bubble spectra.
We follow the notation of Appendix A in Ref.~\cite{Jinno:2017fby}.

\begin{figure}
\begin{center}
\includegraphics[width=0.5\columnwidth]{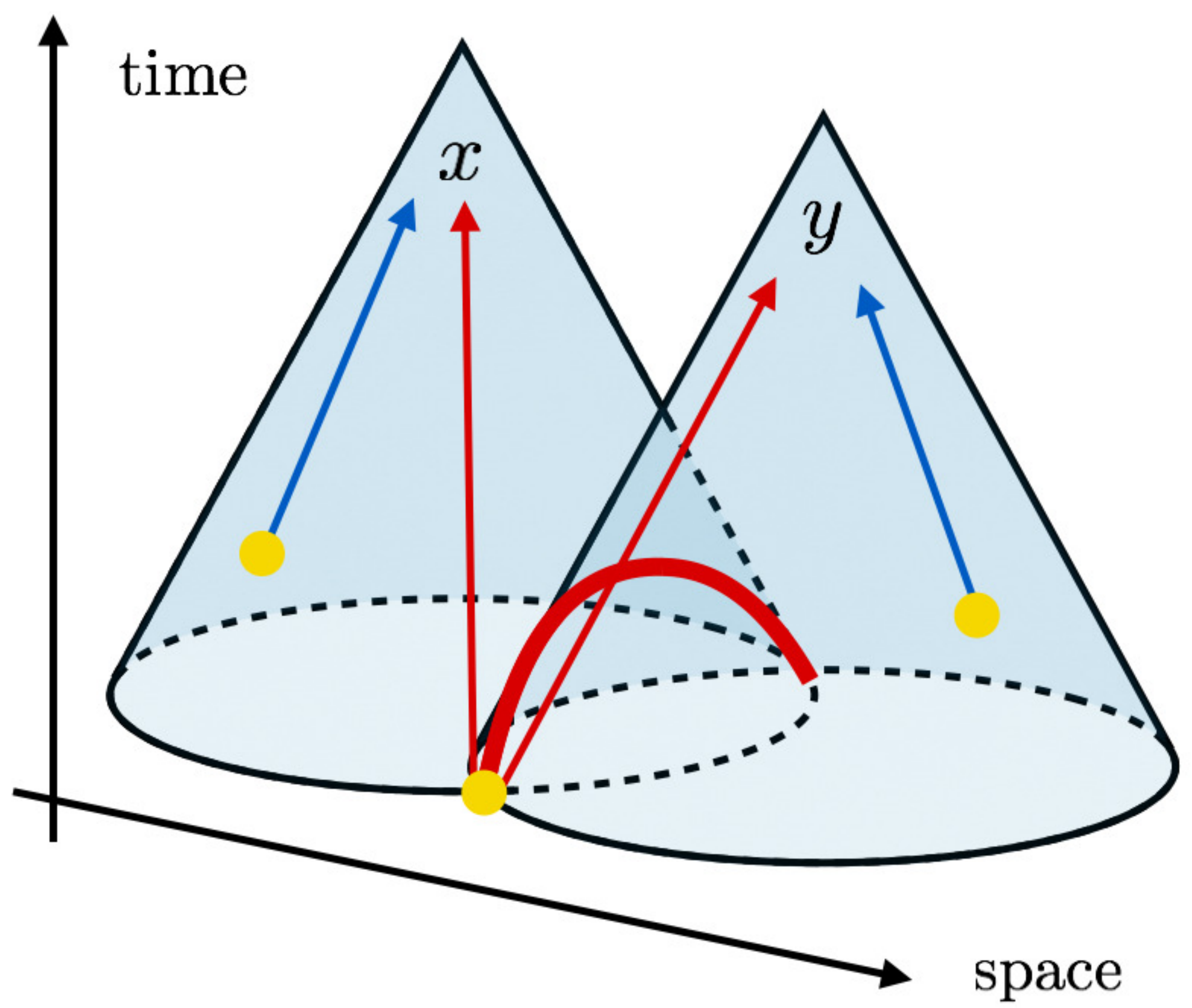} 
\caption{\small
3-dimensional plot of the paths of the bubble wall fragments of $x$ and $y$.
These bubbles nucleate on the surfaces of the past $v$-cones of $x$ and $y$.
Bubble wall propagation in the single- and double-bubble contributions is 
shown as red and blue lines, respectively.
The same figure as in Refs.~\cite{Jinno:2016vai,Jinno:2017fby}.
}
\label{fig:3DLightCone}
\end{center}
\end{figure}

\subsection{False vacuum probability}

Let us consider the probability $P(x,y)$ with which both of the spacetime points $x$ and $y$ 
remain in the false vacuum.
It is equivalent to the probability for no bubbles to nucleate in 
$V_{xy} \equiv V_x \cup V_y$,
where $V_x$ and $V_y$ are the regions inside the past $v$-cones
(past cones with velocity $v$) of $x$ and $y$, respectively
(see Fig.~\ref{fig:3DLightCone}).
Then, the probability is given by
\begin{align}
P(x,y)
&=
e^{-I(x,y)},
\;\;\;\;
I(x,y)
= \int_{V_{xy}} d^4z \; \Gamma(z).
\label{eq:AppFalseP}
\end{align}
Let us write down an explicit form of the function $I$.
The spacial volume of $V_{xy}$ on the constant-time hypersurface $\Sigma_t$ at time $t$
consists of intersecting spheres for $t < t_{\rm max}$,
but separate sphere(s) for $t > t_{\rm max}$,
where $t_{\rm max} \equiv (t_x + t_y - r_v)/2$.
Note that $t_{\rm max}$ is the latest time when two past $v$-cones 
intersect with each other in Fig.~\ref{fig:3DLightCone}. 
Then $I$ is given by
\begin{align}
I(x,y)
&= 
\int_{-\infty}^{t_{\rm max}}dt \; 
\Gamma(t) 
\left[ \frac{\pi}{3} r_x^3(2 + c_{\times x})(1 - c_{\times x})^2 
+ \frac{\pi}{3} r_y^3(2 - c_{\times y})(1 + c_{\times y})^2 \right]
\nonumber \\
&\;\;\;\;
+ 
\int_{t_{\rm max}}^{t_x} dt \; 
\Gamma(t) \frac{4\pi}{3} r_x^3
+
\int_{t_{\rm max}}^{t_y} dt \; 
\Gamma(t) \frac{4\pi}{3} r_y^3.
\label{eq:AppI}
\end{align}
Here $r_x$ and $r_y$ are defined as $r_x \equiv v(t_x - t)$ and $r_y \equiv v(t_y - t)$, respectively, 
and the cosines have relations
\begin{align}
c_{\times x}
&=
- \frac{r_x^2+r^2-r_y^2}{2r_xr},
\;\;\;\;
c_{\times y}
= \frac{r_y^2+r^2-r_x^2}{2r_yr}.
\label{eq:Appcxy}
\end{align}
Note that the quantity in the squared parenthesis in Eq.~(\ref{eq:AppI}) is 
the spacial volume of $V_{xy}$ on $\Sigma_t$ for $t < t_{\rm max}$.
Given the nucleation rate (\ref{eq:AppGamma}), it is straightforward to evaluate this integral. 
We obtain
\begin{align}
I(x,y)
&=
v^3\Gamma_*
\left(
\begin{matrix}
c_{{\rm Exp},t_{x,y}} \\
c_{{\rm Exp},-t_{x,y}} \\
c_{{\rm Exp},-r_v} \\
c_{1 + {\rm Erf},t_{x,y}} \\
c_{1 + {\rm Erf},-t_{x,y}} \\
c_{1 + {\rm Erf},-r_v}
\end{matrix}
\right)^{\rm T}
\renewcommand{\arraystretch}{2.2}
\left(
\begin{matrix}
\displaystyle {\rm Exp}\left[ - \left(t_{\left< x,y \right>} + \frac{t_{x,y}}{2}\right)^2 \right] \\
\displaystyle {\rm Exp}\left[ - \left(t_{\left< x,y \right>} - \frac{t_{x,y}}{2}\right)^2 \right] \\
\displaystyle {\rm Exp}\left[ - \left(t_{\left< x,y \right>} - \frac{r_v}{2}\right)^2 \right] \\
\displaystyle 1 + {\rm Erf}\left[ t_{\left< x,y \right>} + \frac{t_{x,y}}{2} \right] \\
\displaystyle 1 + {\rm Erf}\left[ t_{\left< x,y \right>} - \frac{t_{x,y}}{2} \right] \\
\displaystyle 1 + {\rm Erf}\left[ t_{\left< x,y \right>} - \frac{r_v}{2} \right]
\end{matrix}
\right),
\renewcommand{\arraystretch}{1}
\label{eq:AppI2}
\end{align}
where $t_{\left< x,y \right>} \equiv (t_x + t_y)/2$ and ``T" denotes the transpose.
The coefficients are given by
\begin{align}
&c_{{\rm Exp},t_{x,y}}
=
\frac{\pi}{6}
\left(
\begin{matrix}
1 \\
t_{\left< x,y \right>} \\
t_{\left< x,y \right>}^2
\end{matrix}
\right)^{\rm T}
\left(
\begin{matrix}
4 & 0 & 1 \\
0 & 4 & 0 \\
4 & 0 & 0 
\end{matrix}
\right)
\left(
\begin{matrix}
1 \\
t_{x,y} \\
t_{x,y}^2
\end{matrix}
\right),
\\
&c_{{\rm Exp},-t_{x,y}}
=
\frac{\pi}{6}
\left(
\begin{matrix}
1 \\
t_{\left< x,y \right>} \\
t_{\left< x,y \right>}^2
\end{matrix}
\right)^{\rm T}
\left(
\begin{matrix}
4 & 0 & 1 \\
0 & -4 & 0 \\
4 & 0 & 0 
\end{matrix}
\right)
\left(
\begin{matrix}
1 \\
t_{x,y} \\
t_{x,y}^2
\end{matrix}
\right),
\\
&c_{{\rm Exp},-r_v}
= 
\frac{\pi}{24}
\frac{1}{r_v}
\left(
\begin{matrix}
1 \\
t_{\left< x,y \right>} \\
t_{\left< x,y \right>}^2
\end{matrix}
\right)^{\rm T}
\left(
\begin{matrix}
-16r_v + 2r_v^3 & 0 & -6r_v \\
4r_v^2 & 0 & 12 \\
-16r_v & 0 & 0
\end{matrix}
\right)
\left(
\begin{matrix}
1 \\
t_{x,y} \\
t_{x,y}^2
\end{matrix}
\right),
\end{align}
\begin{align}
&c_{1 + {\rm Erf},t_{x,y}}
=
\frac{\pi^{3/2}}{12}
\left(
\begin{matrix}
1 \\
t_{\left< x,y \right>} \\
t_{\left< x,y \right>}^2 \\
t_{\left< x,y \right>}^3
\end{matrix}
\right)^{\rm T}
\left(
\begin{matrix}
0 & 6 & 0 & 1 \\
12 & 0 & 6 & 0 \\
0 & 12 & 0 & 0 \\
8 & 0 & 0 & 0
\end{matrix}
\right)
\left(
\begin{matrix}
1 \\
t_{x,y} \\
t_{x,y}^2 \\
t_{x,y}^3
\end{matrix}
\right),
\\
&c_{1 + {\rm Erf},-t_{x,y}}
=
\frac{\pi^{3/2}}{12}
\left(
\begin{matrix}
1 \\
t_{\left< x,y \right>} \\
t_{\left< x,y \right>}^2 \\
t_{\left< x,y \right>}^3
\end{matrix}
\right)^{\rm T}
\left(
\begin{matrix}
0 & -6 & 0 & -1 \\
12 & 0 & 6 & 0 \\
0 & -12 & 0 & 0 \\
8 & 0 & 0 & 0
\end{matrix}
\right)
\left(
\begin{matrix}
1 \\
t_{x,y} \\
t_{x,y}^2 \\
t_{x,y}^3
\end{matrix}
\right),
\\
&c_{1 + {\rm Erf},-r_v}
= 
\frac{\pi^{3/2}}{24}
\frac{1}{r_v}
\left(
\begin{matrix}
1 \\
t_{\left< x,y \right>} \\
t_{\left< x,y \right>}^2 \\
t_{\left< x,y \right>}^3
\end{matrix}
\right)^{\rm T}
\left(
\begin{matrix}
6r_v^2 - r_v^4 & 0 & 6 + 3r_v^2 \\
-24r_v & 0 & -12r_v \\
12r_v^2 & 0 & 12 \\
-16r_v & 0 & 0
\end{matrix}
\right)
\left(
\begin{matrix}
1 \\
t_{x,y} \\
t_{x,y}^2
\end{matrix}
\right).
\end{align}

\subsection{Single-bubble spectrum}

Let us now derive the single-bubble spectrum.
In the thin-wall limit,
the necessary and sufficient conditions 
for the wall of a single bubble to contribute to the energy-momentum tensor 
at both of the spacetime points $x$ and $y$ are summarized as
\begin{itemize}
\item
No bubble nucleates in $V_{xy}$.

\item
One bubble nucleates in $\delta V_{xy}$.
\end{itemize}
Here $\delta V_{xy}$, shown as the red line in Fig.~\ref{fig:3DLightCone}, 
is the narrow four-dimensional region on the two past $v$-cones:
$\delta V_{xy} \equiv (V_{x + \delta} - V_x) \cap (V_{y + \delta} - V_y)$
with $x + \delta \equiv (t_x + l_B/v,\vec{x})$ and $y + \delta \equiv (t_y + l_B/v,\vec{y})$.
Following the same procedure as in Appendix A in Ref.~\cite{Jinno:2017fby}, we obtain
\begin{align}
&\Pi^{(s)}(t_x,t_y,k)
\nonumber \\
&= 
\frac{4\pi^2}{9}\kappa^2\rho_0^2 
\int_{v|t_{x,y}|}^\infty dr 
\int_{-\infty}^{t_{\rm max}}dt_n \;
e^{-I(x,y)}
\Gamma(t_n) \;
r r_x^{(s)2}r_y^{(s)2}
\left[
j_0(kr){\mathcal K}_{\times 0} 
+ \frac{j_1(kr)}{kr}{\mathcal K}_{\times 1} 
+ \frac{j_2(kr)}{(kr)^2}{\mathcal K}_{\times 2}
\right].
\label{eq:AppSPi}
\end{align}
Here $t_n$ denotes the nucleation time of the bubble nucleated in $\delta V_{xy}$.
See Ref.~\cite{Jinno:2017fby} for the definition of the other quantities.
In Eq.~(\ref{eq:AppSPi}), all the $l_B$'s have been canceled out 
because the allowed volume for bubble nucleation (the thick red line in Fig.~\ref{fig:3DLightCone})
is proportional to $l_B^2$ while the resulting $T(x)T(y)$ is proportional to $l_B^{-2}$.
Also, $j_0$, $j_1$ and $j_2$ are the spherical Bessel functions defined as
\begin{align}
j_0(x)
&= \frac{\sin x}{x},
\;\;\;\;
j_1(x)
= \frac{\sin x - x \cos x}{x^2},
\;\;\;\;
j_2(x)
= \frac{(3 - x^2)\sin x - 3x \cos x}{x^3}.
\label{eq:Appj}
\end{align}
After integrating out the nucleation time, we obtain
\begin{align}
\Pi^{(s)}(t_x,t_y,k)
&= 
\frac{4\pi^2}{3} v^6\kappa^2\rho_0^2 \;
\Gamma_*
\int_{|t_{x,y}|}^\infty dr_v \;
e^{-I(x,y)}
\left[
j_0(vkr_v){\mathcal S}_0 
+ \frac{j_1(vkr_v)}{vkr_v}{\mathcal S}_1 
+ \frac{j_2(vkr_v)}{(vkr_v)^2}{\mathcal S}_2
\right],
\end{align}
where ${\mathcal S}_0$, ${\mathcal S}_1$ and ${\mathcal S}_2$ are given by
\begin{align}
{\mathcal S}_0
&= 
c_{{\rm Exp},0}^{(s)} {\rm Exp}\left[ - \left( t_{\left< x,y \right>} - \frac{r_v}{2} \right)^2 \right] 
+
c_{1 + {\rm Erf},0}^{(s)} \left( 1 + {\rm Erf}\left[ t_{\left< x,y \right>} - \frac{r_v}{2} \right] \right),
\nonumber \\
{\mathcal S}_1
&= 
c_{{\rm Exp},1}^{(s)} {\rm Exp}\left[ - \left( t_{\left< x,y \right>} - \frac{r_v}{2} \right)^2 \right] 
+
c_{1 + {\rm Erf},1}^{(s)} \left( 1 + {\rm Erf}\left[ t_{\left< x,y \right>} - \frac{r_v}{2} \right] \right),
\nonumber \\
{\mathcal S}_2
&= 
c_{{\rm Exp},2}^{(s)} {\rm Exp}\left[ - \left( t_{\left< x,y \right>} - \frac{r_v}{2} \right)^2 \right] 
+
c_{1 + {\rm Erf},2}^{(s)} \left( 1 + {\rm Erf}\left[ t_{\left< x,y \right>} - \frac{r_v}{2} \right] \right),
\label{eq:AppS}
\end{align}
with the coefficients
\begin{align}
&c_{{\rm Exp},0}^{(s)}
= 
\frac{1}{96}
\frac{(t_{x,y}^2 - r_v^2)^2}{r_v^3}
\left(
\begin{matrix}
1 \\
t_{\left< x,y \right>} \\
t_{\left< x,y \right>}^2 \\
t_{\left< x,y \right>}^3
\end{matrix}
\right)^{\rm T}
\left(
\begin{matrix}
12 r_v - 2r_v^3 \\
40 - 4 r_v^2 \\
8 r_v \\
16
\end{matrix}
\right),
\\
&c_{1 + {\rm Erf},0}^{(s)}
= 
\frac{\sqrt{\pi}}{96}
\frac{(t_{x,y}^2 - r_v^2)^2}{r_v^3}
\left(
\begin{matrix}
1 \\
t_{\left< x,y \right>} \\
t_{\left< x,y \right>}^2 \\
t_{\left< x,y \right>}^3 \\
t_{\left< x,y \right>}^4
\end{matrix}
\right)^{\rm T}
\left(
\begin{matrix}
12 - 4 r_v^2 + r_v^4 \\
0 \\
48 - 8 r_v^2 \\
0 \\
16
\end{matrix}
\right),
\end{align}
\begin{align}
&c_{{\rm Exp},1}^{(s)}
= 
\frac{1}{96}
\frac{t_{x,y}^2 - r_v^2}{r_v^3}
\left(
\begin{matrix}
1 \\
t_{\left< x,y \right>} \\
t_{\left< x,y \right>}^2 \\
t_{\left< x,y \right>}^3
\end{matrix}
\right)^{\rm T}
\left(
\begin{matrix}
24 r_v^3 + 12 r_v^5 & 0 & -120r_v + 4 r_v^3 \\
80 r_v^2 + 24 r_v^4 & 0 & -400 + 8 r_v^2 \\
16 r_v^3 & 0 & -80 r_v \\
32 r_v^2 & 0 & -160
\end{matrix}
\right)
\left(
\begin{matrix}
1 \\
t_{x,y} \\
t_{x,y}^2
\end{matrix}
\right),
\\
&c_{1 + {\rm Erf},1}^{(s)}
= 
\frac{\sqrt{\pi}}{96}
\frac{t_{x,y}^2 - r_v^2}{r_v^3}
\left(
\begin{matrix}
1 \\
t_{\left< x,y \right>} \\
t_{\left< x,y \right>}^2 \\
t_{\left< x,y \right>}^3 \\
t_{\left< x,y \right>}^4
\end{matrix}
\right)^{\rm T}
\left(
\begin{matrix}
24 r_v^2 + 8 r_v^4 - 6 r_v^6 & 0 & -120 + 24 r_v^2 - 2 r_v^4 \\
0 & 0 & 0 \\
96 r_v^2 + 16 r_v^4 & 0 & -480 + 48 r_v^2 \\
0 & 0 & 0 \\
32 r_v^2 & 0 & -160 
\end{matrix}
\right)
\left(
\begin{matrix}
1 \\
t_{x,y} \\
t_{x,y}^2
\end{matrix}
\right),
\end{align}
\begin{align}
&c_{{\rm Exp},2}^{(s)}
= 
\frac{1}{96}
\frac{1}{r_v^3}
\left(
\begin{matrix}
1 \\
t_{\left< x,y \right>} \\
t_{\left< x,y \right>}^2 \\
t_{\left< x,y \right>}^3
\end{matrix}
\right)^{\rm T}
\left(
\begin{matrix}
36 r_v^5 + 10 r_v^7 & 0 & -360 r_v^3 - 36 r_v^5 & 0 & 420r_v + 10 r_v^3 \\
120 r_v^4 + 20 r_v^6 & 0 & -1200 r_v^2 - 72 r_v^4 & 0 & 1400 + 20 r_v^2 \\
24 r_v^5 & 0 & -240 r_v^3 & 0 & 280 r_v \\
48 r_v^4 & 0 & -480 r_v^2 & 0 & 560
\end{matrix}
\right)
\left(
\begin{matrix}
1 \\
t_{x,y} \\
t_{x,y}^2 \\
t_{x,y}^3 \\
t_{x,y}^4
\end{matrix}
\right),
\\
&c_{1 + {\rm Erf},2}^{(s)}
= 
\frac{\sqrt{\pi}}{96}
\frac{1}{r_v^3}
\nonumber \\
&\times
\left(
\begin{matrix}
1 \\
t_{\left< x,y \right>} \\
t_{\left< x,y \right>}^2 \\
t_{\left< x,y \right>}^3 \\
t_{\left< x,y \right>}^4
\end{matrix}
\right)^{\rm T}
\left(
\begin{matrix}
36 r_v^4 + 4 r_v^6 + 3 r_v^8 & 0 & -360 r_v^2 + 24 r_v^4 + 2 r_v^6 & 0 & 420 - 60 r_v^2 + 3 r_v^4 \\
0 & 0 & 0 & 0 & 0 \\
144 r_v^4 + 8 r_v^6 & 0 & -1440 r_v^2 + 48 r_v^4 & 0 & 1680 - 120 r_v^2 \\
0 & 0 & 0 & 0 & 0 \\
48 r_v^4 & 0 & -480 r_v^2 & 0 & 560
\end{matrix}
\right)
\left(
\begin{matrix}
1 \\
t_{x,y} \\
t_{x,y}^2 \\
t_{x,y}^3 \\
t_{x,y}^4
\end{matrix}
\right).
\end{align}
Then, the single-bubble spectrum is obtained by using Eq.~(\ref{eq:DeltaPi}):
\begin{align}
\Delta^{(s)}
= 
\beta^2 v^6 k^3 \;
\Gamma_*
&\int_{-\infty}^\infty dt_{\left< x,y \right>}
\int_{-\infty}^\infty dt_{x,y}
\int_{|t_{x,y}|}^\infty dr_v 
\nonumber \\
&e^{-I(x,y)}
\left[
j_0(vkr_v){\mathcal S}_0
+ \frac{j_1(vkr_v)}{vkr_v}{\mathcal S}_1 
+ \frac{j_2(vkr_v)}{(vkr_v)^2}{\mathcal S}_2
\right]
\cos(kt_{x,y}).
\label{eq:AppDeltaS}
\end{align}

\subsection{Double-bubble spectrum}

Next let us derive the double-bubble spectrum.
The necessary and sufficient conditions for the walls of two different bubbles to contribute to 
the energy-momentum tensor at $x$ and $y$ are summarized as
\begin{itemize}
\item
No bubble nucleates in $V_{xy}$.

\item
One bubble nucleates in $\delta V_x^{(y)}$,
and another nucleates in $\delta V_y^{(x)}$.
\end{itemize}
Here $\delta V_x^{(y)}$ and $\delta V_y^{(x)}$ are thin surfaces of $V_x$ and $V_y$ in Fig.~\ref{fig:3DLightCone}: 
$\delta V_x^{(y)} \equiv (V_{x + \delta} - V_x) - V_{y + \delta}$
and $\delta V_y^{(x)} \equiv (V_{y + \delta} - V_y) - V_{x + \delta}$.
Following the same procedure as in Appendix A in Ref.~\cite{Jinno:2017fby}, we obtain
\begin{align}
\Pi^{(d)}(t_x,t_y,k)
= 
\frac{16\pi^3}{9}\kappa^2\rho_0^2 
&\int_{v|t_{x,y}|}^\infty dr 
\int_{-\infty}^{t_{\rm max}}dt_{xn}
\int_{-\infty}^{t_{\rm max}}dt_{yn} 
\nonumber \\
&
e^{-I(x,y)}
\Gamma(t_{xn})\Gamma(t_{yn}) \;
r^2 r_x^{(d)3}r_y^{(d)3}
(c_{\times x} - c_{\times x}^3)(c_{\times y}^3 - c_{\times y})
\frac{j_2(kr)}{(kr)^2}.
\label{eq:AppDPi}
\end{align}
Here $t_{xn}$ and $t_{yn}$ are the nucleation times of the two bubbles
nucleated in $\delta V_x^{(y)}$ and $\delta V_y^{(x)}$, respectively.
In Eq.~(\ref{eq:AppDPi}), all the $l_B$'s have been canceled out 
because the allowed volume for bubble nucleation (the thin surface of past cones in Fig.~\ref{fig:3DLightCone})
is proportional to $l_B$ for each bubble while the resulting $T(x)T(y)$ is proportional to $l_B^{-2}$.
These time integrations can be performed explicitly, and we obtain
\begin{align}
\Pi^{(d)}(t_x,t_y,k)
&= 
\frac{4\pi^2}{3} v^9\kappa^2\rho_0^2 \;
\Gamma_*^2 \;
\int_{|t_{x,y}|}^\infty dr_v \;
e^{-I(x,y)}
\left[
\frac{j_2(vkr_v)}{(vkr_v)^2}
{\mathcal D}_2
\right],
\end{align}
where ${\mathcal D}_2$ is given by
\begin{align}
{\mathcal D}_2
&=
\left[
c_{{\rm Exp},2}^{(d)}(t_{x,y}) {\rm Exp}\left[ - \left( t_{\left< x,y \right>} - \frac{r_v}{2} \right)^2 \right] 
+
c_{1 + {\rm Erf},2}^{(d)}(t_{x,y}) \left( 1 + {\rm Erf}\left[ t_{\left< x,y \right>} - \frac{r_v}{2} \right] \right)
\right]
\nonumber \\
&\;\;\;\;\;
\times
\left[
c_{{\rm Exp},2}^{(d)}(-t_{x,y}) {\rm Exp}\left[ - \left( t_{\left< x,y \right>} - \frac{r_v}{2} \right)^2 \right] 
+
c_{1 + {\rm Erf},2}^{(d)}(-t_{x,y}) \left( 1 + {\rm Erf}\left[ t_{\left< x,y \right>} - \frac{r_v}{2} \right] \right)
\right]
\label{eq:AppD}
\end{align}
with the coefficients
\begin{align}
&c_{{\rm Exp},2}^{(d)}(t_{x,y})
= 
\sqrt{\frac{\pi}{192}}
\frac{t_{x,y}^2 - r_v^2}{r_v^2}
\left(
\begin{matrix}
1 \\
t_{\left< x,y \right>} \\
t_{\left< x,y \right>}^2
\end{matrix}
\right)^{\rm T}
\left(
\begin{matrix}
2 r_v^3 & 8 \\
4 r_v^2 & 4r_v \\
0 & 8
\end{matrix}
\right)
\left(
\begin{matrix}
1 \\
t_{x,y}
\end{matrix}
\right),
\\
&c_{1 + {\rm Erf},2}^{(d)}(t_{x,y})
= 
\frac{\pi}{\sqrt{192}}
\frac{t_{x,y}^2 - r_v^2}{r_v^2}
\left(
\begin{matrix}
1 \\
t_{\left< x,y \right>} \\
t_{\left< x,y \right>}^2 \\
t_{\left< x,y \right>}^3
\end{matrix}
\right)^{\rm T}
\left(
\begin{matrix}
2r_v^2 - r_v^4 & 0 \\
0 & 12 - 2 r_v^2 \\
4r_v^2 & 0 \\
0 & 8
\end{matrix}
\right)
\left(
\begin{matrix}
1 \\
t_{x,y}
\end{matrix}
\right).
\end{align}
Then, the double-bubble spectrum is obtained by using Eq.~(\ref{eq:DeltaPi}):
\begin{align}
\Delta^{(d)}
&= 
\beta^2 v^9 k^3 \;
\Gamma_*^2
\int_{-\infty}^\infty dt_{\left< x,y \right>}
\int_{-\infty}^\infty dt_{x,y}
\int_{|t_{x,y}|}^\infty dr_v \;
e^{-I(x,y)}
\left[
\frac{j_2(vkr_v)}{(vkr_v)^2}{\mathcal D}_2
\right]
\cos(kt_{x,y}).
\label{eq:AppDeltaD}
\end{align}

\subsection{Spectrum with $\delta$-function nucleation rate} 
\label{app:AppDelta}

In this subsection we present the spectrum with $\delta$-function nucleation rate.
We parameterize the nucleation rate as
\begin{align}
\Gamma(t)
&= n_* \delta (t).
\label{eq:AppGammadelta}
\end{align}
Note that the spectral shape for $\gamma \to \infty$ with the Gaussian nucleation rate (\ref{eq:Gamma})
(or equivalently Eq.~(\ref{eq:Gammapr}) or (\ref{eq:AppGamma}))
approaches the one with this nucleation rate.
This is understood as follows.
As mentioned in Sec.~\ref{subsec:AA}, one may eliminate one parameter 
from the original nucleation rate (\ref{eq:Gamma}).
If we use Parameterization 2 for example (see Table~\ref{tbl:Param}),
we have free parameters $(\beta', \gamma)$.
In large $\gamma/\beta'$ limit, 
the typical time interval for bubble nucleation is given by $1/\gamma$
since the nucleation rate is Gaussian,
while the number density for bubble nucleation points roughly becomes
$\int dt \; \Gamma(t) \sim \beta'^4/\gamma$.
The latter means that the typical distance for neighboring bubbles scale as $\gamma^{1/3}$.
Therefore, in large $\gamma/\beta'$ limit, the dispersion in the bubble nucleation time 
becomes negligible compared to the timescale of bubble expansion and collisions,
and the resulting GW spectrum approaches the one with the nucleation rate (\ref{eq:AppGammadelta}).
A similar argument applies to Parameterization 3.

Below we first give the expression for the false-vacuum probability, 
and then show the expressions for the GW spectrum.

\subsubsection*{False-vacuum probability}

The expressions (\ref{eq:AppFalseP})--(\ref{eq:Appcxy}) are the same as the Gaussian case.
Substituting Eq.~(\ref{eq:AppGammadelta}) into these expressions, we obtain
\begin{align}
\frac{I}{v^3n_*}
&= 
\Theta \left( - t_{\left< x,y \right>} + \frac{r_v}{2} \right) \;
\Theta \left( t_{\left< x,y \right>} + \frac{t_{x,y}}{2} \right) 
\frac{\pi}{3}
\left(
\begin{matrix}
1 \\
t_{\left< x,y \right>} \\
t_{\left< x,y \right>}^2 \\
t_{\left< x,y \right>}^3
\end{matrix}
\right)^{\rm T}
\left(
\begin{matrix}
0 & 0 & 0 & 1 \\
0 & 0 & 6 & 0 \\
0 & 12 & 0 & 0 \\
8 & 0 & 0 & 0
\end{matrix}
\right)
\left(
\begin{matrix}
1 \\
t_{x,y} \\
t_{x,y}^2 \\
t_{x,y}^3
\end{matrix}
\right),
\nonumber \\
&\;\;\;\;
+
\Theta \left( - t_{\left< x,y \right>} + \frac{r_v}{2} \right) \;
\Theta \left( t_{\left< x,y \right>} - \frac{t_{x,y}}{2} \right) 
\frac{\pi}{3}
\left(
\begin{matrix}
1 \\
t_{\left< x,y \right>} \\
t_{\left< x,y \right>}^2 \\
t_{\left< x,y \right>}^3
\end{matrix}
\right)^{\rm T}
\left(
\begin{matrix}
0 & 0 & 0 & -1 \\
0 & 0 & 6 & 0 \\
0 & -12 & 0 & 0 \\
8 & 0 & 0 & 0
\end{matrix}
\right)
\left(
\begin{matrix}
1 \\
t_{x,y} \\
t_{x,y}^2 \\
t_{x,y}^3
\end{matrix}
\right),
\nonumber \\
&\;\;\;\;
+ 
\Theta \left( t_{\left< x,y \right>} - \frac{r_v}{2} \right) \;
\frac{\pi}{12}
\frac{1}{r_v}
\left(
\begin{matrix}
1 \\
t_{\left< x,y \right>} \\
t_{\left< x,y \right>}^2 \\
t_{\left< x,y \right>}^3
\end{matrix}
\right)^{\rm T}
\left(
\begin{matrix}
- r_v^4 & 0 & 3r_v^2 \\
0 & 0 & 12r_v \\
12r_v^2 & 0 & 12 \\
16r_v & 0 & 0
\end{matrix}
\right)
\left(
\begin{matrix}
1 \\
t_{x,y} \\
t_{x,y}^2
\end{matrix}
\right).
\label{eq:AppI3}
\end{align}
Note that this can also be derived from Eq.~(\ref{eq:AppI2}).
This is because Gaussian nucleation rate (\ref{eq:Gamma}) with $\gamma \to \infty$
should coincide with $\delta$-function nucleation rate.
In fact, restoring $\gamma$ and then taking $\gamma \to \infty$ limit 
one can check that Eq.~(\ref{eq:AppI2}) reduces to Eq.~(\ref{eq:AppI3}).\footnote{
The limit $\gamma \to \infty$ corresponds to the replacement 
${\rm Exp}[\dots] \to 0$ and $1 + {\rm Erf}[\dots] \to 2 \Theta (\dots)$
in Eq.~(\ref{eq:AppI2}).
After this, use
\begin{align}
\Theta \left( t_{\left< x,y \right>} + \frac{t_{x,y}}{2} \right)
&= 
\Theta \left( t_x \right) \Theta \left( t_{\left< x,y \right>} - \frac{r_v}{2} \right)
+ \Theta \left( t_x \right) \Theta \left( - t_{\left< x,y \right>} + \frac{r_v}{2} \right),
\label{eq:Apptheta1}
\\
\Theta \left( t_{\left< x,y \right>} - \frac{t_{x,y}}{2} \right)
&= 
\Theta \left( t_y \right) \Theta \left( t_{\left< x,y \right>} - \frac{r_v}{2} \right)
+ \Theta \left( t_y \right) \Theta \left( - t_{\left< x,y \right>} + \frac{r_v}{2} \right).
\label{eq:Apptheta2}
\end{align}
In the R.H.S.s, $\Theta(t_x)$ and $\Theta(t_y)$ in each first terms can be set to unity 
because $t_{\left< x,y \right>} - r_v/2 > 0$ and $|t_{x,y}| - r_v < 0$
(the latter coming from the fact that we adopt the envelope approximation)
guarantee $t_x > 0$ and $t_y > 0$.
Then, after arranging and identifying $\Gamma_*$ with $n_*$, one obtains Eq.~(\ref{eq:AppI3}). 
}

\subsubsection*{Single-bubble spectrum}

The procedure to obtain the spectrum is the same as in the Gaussian case.
As a result, we obtain Eq.~(\ref{eq:AppDeltaS}) with $\Gamma_*$ replaced by $n_*$ 
and ${\mathcal S}$ functions given by
\begin{align}
&{\mathcal S}_0
= 
\Theta \left( t_{\left< x,y \right>} - \frac{r_v}{2} \right) \;
\frac{1}{48}
\frac{(t_{x,y}^2 - r_v^2)^2}{r_v^3}
\left(
\begin{matrix}
1 \\
t_{\left< x,y \right>} \\
t_{\left< x,y \right>}^2 \\
t_{\left< x,y \right>}^3 \\
t_{\left< x,y \right>}^4
\end{matrix}
\right)^{\rm T}
\left(
\begin{matrix}
r_v^4 \\
0 \\
- 8 r_v^2 \\
0 \\
16
\end{matrix}
\right),
\\
&{\mathcal S}_1
= 
\Theta \left( t_{\left< x,y \right>} - \frac{r_v}{2} \right) \;
\frac{1}{48}
\frac{t_{x,y}^2 - r_v^2}{r_v^3}
\left(
\begin{matrix}
1 \\
t_{\left< x,y \right>} \\
t_{\left< x,y \right>}^2 \\
t_{\left< x,y \right>}^3 \\
t_{\left< x,y \right>}^4
\end{matrix}
\right)^{\rm T}
\left(
\begin{matrix}
- 6 r_v^6 & 0 & - 2 r_v^4 \\
0 & 0 & 0 \\
16 r_v^4 & 0 & 48 r_v^2 \\
0 & 0 & 0 \\
32 r_v^2 & 0 & -160 
\end{matrix}
\right)
\left(
\begin{matrix}
1 \\
t_{x,y} \\
t_{x,y}^2
\end{matrix}
\right),
\\
&{\mathcal S}_2
= 
\Theta \left( t_{\left< x,y \right>} - \frac{r_v}{2} \right) \;
\frac{1}{48}
\frac{1}{r_v^3}
\left(
\begin{matrix}
1 \\
t_{\left< x,y \right>} \\
t_{\left< x,y \right>}^2 \\
t_{\left< x,y \right>}^3 \\
t_{\left< x,y \right>}^4
\end{matrix}
\right)^{\rm T}
\left(
\begin{matrix}
3 r_v^8 & 0 & 2 r_v^6 & 0 & 3 r_v^4 \\
0 & 0 & 0 & 0 & 0 \\
8 r_v^6 & 0 & 48 r_v^4 & 0 & - 120 r_v^2 \\
0 & 0 & 0 & 0 & 0 \\
48 r_v^4 & 0 & -480 r_v^2 & 0 & 560
\end{matrix}
\right)
\left(
\begin{matrix}
1 \\
t_{x,y} \\
t_{x,y}^2 \\
t_{x,y}^3 \\
t_{x,y}^4
\end{matrix}
\right).
\end{align}
This spectrum can also be obtained by restoring $\gamma$ and taking $\gamma \to \infty$ limit
in Eq.~(\ref{eq:AppDeltaS}).

\subsubsection*{Double-bubble spectrum}

The procedure is also the same as in the Gaussian case.
As a result, we obtain Eq.~(\ref{eq:AppDeltaD}) with $\Gamma_*$ replaced by $n_*$ 
and ${\mathcal D}$ function given by
\begin{align}
{\mathcal D}_2
&= 
\left[
\sqrt{\frac{\pi}{48}}
\frac{t_{x,y}^2 - r_v^2}{r_v^2}
\left(
\begin{matrix}
1 \\
t_{\left< x,y \right>} \\
t_{\left< x,y \right>}^2 \\
t_{\left< x,y \right>}^3
\end{matrix}
\right)^{\rm T}
\left(
\begin{matrix}
- r_v^4 & 0 \\
0 & - 2 r_v^2 \\
4r_v^2 & 0 \\
0 & 8
\end{matrix}
\right)
\left(
\begin{matrix}
1 \\
t_{x,y}
\end{matrix}
\right)
\right]
\nonumber \\
&\;\;\;\;\;\;\;\;\;\;
\times
\left[
\sqrt{\frac{\pi}{48}}
\frac{t_{x,y}^2 - r_v^2}{r_v^2}
\left(
\begin{matrix}
1 \\
t_{\left< x,y \right>} \\
t_{\left< x,y \right>}^2 \\
t_{\left< x,y \right>}^3
\end{matrix}
\right)^{\rm T}
\left(
\begin{matrix}
- r_v^4 & 0 \\
0 & - 2 r_v^2 \\
4r_v^2 & 0 \\
0 & 8
\end{matrix}
\right)
\left(
\begin{matrix}
1 \\
- t_{x,y}
\end{matrix}
\right)
\right].
\end{align}
This spectrum can also be obtained by restoring $\gamma$ and taking $\gamma \to \infty$ limit
in Eq.~(\ref{eq:AppDeltaD}).

\section{Asymptotic behavior of the spectrum for small Gaussian corrections}
\label{app:Asymptotic}
\setcounter{equation}{0}

In Fig.~\ref{fig:Distinguish} in Sec.~\ref{sec:Numerical}, 
we extrapolated the deviation in the spectral shape for small $\gamma$ 
due to numerical difficulties arising in this limit.
In this appendix we check the validity of this extrapolation
by showing the asymptotic behavior of the spectrum for small $\gamma$.

\begin{figure}[ht]
\begin{center}
\includegraphics[width=0.6\columnwidth]{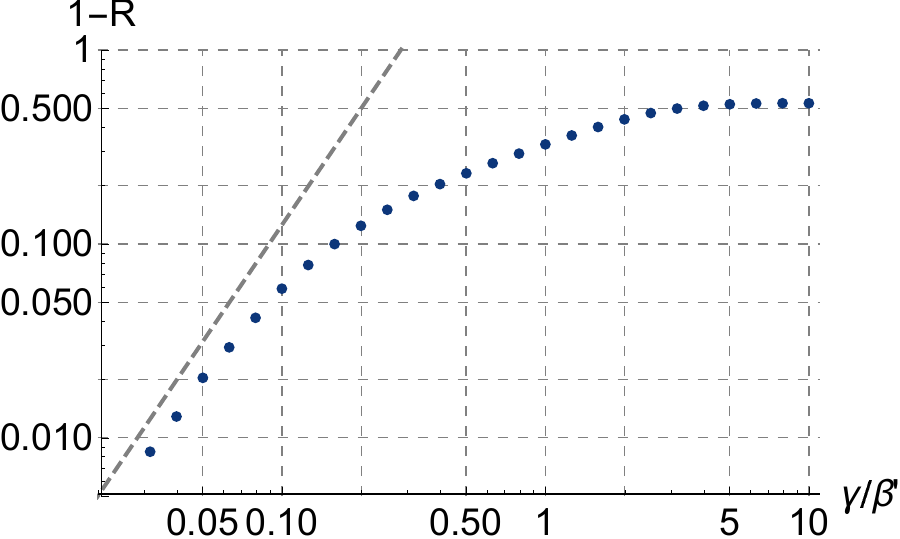} 
\caption{\small
Plot for the deviation $1 - R$ of the spectrum $\tilde{\Delta}$ from the one with $\gamma = 0$,
evaluated at $\tilde{k} = 0.01$. 
The dashed line is proportional to $(\gamma/\beta')^2$.
It is shown that the data points behave $\propto (\gamma/\beta')^2$ for small $\gamma/\beta'$.
}
\label{fig:Asymptotic}
\end{center}
\end{figure}

In fig.~\ref{fig:Asymptotic} 
we plot the deviation $1 - R$ of the normalized spectrum $\tilde{\Delta}$
for various values of $\gamma/\beta'$ at $\tilde{k} = 0.01$.
Here $R$ and its argument $\tilde{k}$ are defined in Eqs.~(\ref{eq:Deltatilde}) and (\ref{eq:R}).
As mentioned in Sec.~\ref{sec:Numerical}, 
the deviation of the spectral shape from the one with $\gamma = 0$ 
is expected to be proportional to $(\gamma/\beta')^2$ for small $\gamma/\beta'$,
because the Gaussian correction appears in the nucleation rate in the form of $\gamma^2$
and, in addition, $\gamma/\beta'$ is the only parameter which determines the spectral shape
(see Eq.~(\ref{eq:Gammapr})).
As seen from Fig.~\ref{fig:Asymptotic},
the deviation $1 - R$ indeed behaves proportional to $\gamma^2/\beta'^2$ for small $\gamma/\beta'$.

\vskip 0.5in

\clearpage

\small
\bibliography{ref}

\end{document}